\documentclass[pre,showpacs,twocolumn,preprintnumbers,amsmath,amssymb]{revtex4}

\usepackage{graphicx}
\usepackage{dcolumn}
\usepackage{bm}
\usepackage{parskip}
\usepackage{subfigure}
\usepackage{amssymb}
\usepackage{amsmath}
\usepackage{amssymb}
\usepackage{graphicx}
\newcommand{\bare}[1]{\mathaccent"7017{#1}}
\def\be{\begin{eqnarray}}
\def\ee{\end{eqnarray}}
\def\be{\begin{equation}}
\def\ee{\end{equation}}
\begin{document}
\title{%
Critical free energy and
Casimir forces
  in
rectangular geometries}

\author{Volker Dohm}

\affiliation{Institute for Theoretical Physics, RWTH Aachen
University, D-52056 Aachen, Germany}

\date {7 June 2011 }

\begin{abstract}
We study the critical behavior of the free energy and the thermodynamic Casimir force in a $L_\parallel^{d-1} \times L$ block geometry in $2<d<4$ dimensions with aspect ratio $\rho=L/L_\parallel$  on the basis of the O$(n)$ symmetric $\varphi^4$ lattice model with periodic boundary conditions and with isotropic short-range interactions. Exact results are derived in the large - $n$ limit describing the geometric crossover from film ($\rho =0$) over cubic ($\rho=1$) to cylindrical ($\rho = \infty$) geometries. For $n=1$, three perturbation approaches in the minimal renormalization scheme at fixed $d$ are presented that cover both the central finite-size regime near $T_c$ for  $1/4 \lesssim \rho \lesssim 3$ and the region well above and below $T_c$. At bulk $T_c$  we predict the critical Casimir force in the vertical $(L)$ direction to be negative (attractive) for a slab ($\rho < 1$), positive (repulsive) for a rod ($\rho > 1$), and zero for a cube $(\rho=1)$. Our results for finite-size scaling functions  agree well with Monte Carlo data for the three-dimensional Ising model by Hasenbusch for $\rho=1$ and by Vasilyev et al. for $\rho=1/6$ above, at, and below $T_c$.

\end{abstract}
\maketitle

\renewcommand{\thesection}{\Roman{section}}
\renewcommand{\theequation}{1.\arabic{equation}}
\setcounter{equation}{0}
\section*{I. Introduction and overview}

In the theory of finite-size effects near phase transitions, the study of critical Casimir forces \cite{fisher78} has remained on a highly active level over the past two decades \cite{krech}. Close to criticality and for sufficiently large confining lengths, such forces are predicted to exhibit universal features in the sense that, for {\it isotropic} systems with short-range interactions, their scaling functions depend only on the geometric shape, on the boundary conditions (b.c.), and on the universality class of the system. For {\it anisotropic} systems of the same universality class (e.g., lattice systems with noncubic symmetry such as anisotropic superconductors \cite{wil-1}), however, two-scale factor universality \cite{priv} is absent
\cite{cd2004,dohm2006,dohm2008,kastening-dohm,DG,newref} and the critical Casimir forces are nonuniversal as they depend on the lattice structure and on the microscopic couplings through a matrix of nonuniversal anisotropy parameters. This implies that the Casimir amplitudes  at bulk $T_c$ depend, in general, on $d(d+1)/2 - 1$ nonuniversal parameters in $d$ dimensional anisotropic systems with given shape and given b.c. \cite{cd2004,dohm2006,dohm2008,kastening-dohm,DG}. This prediction is readily testable by Monte Carlo (MC) simulations of anisotropic spin models; it was also noted in \cite{dohm2008} that experiments in anisotropic superconducting films  \cite{wil-1} could, in principle, demonstrate the nonuniversality of the critical Casimir force \cite{real}. A corresponding nonuniversality of the Binder cumulant \cite{priv,bin-2} has been predicted \cite{cd2004,dohm2008} and has been confirmed by MC simulations  for the anisotropic Ising model \cite{selke2005}.

In our present paper, the focus is on isotropic systems with periodic b.c. for the ($n=1$) Ising universality class on the basis of  the O$(n)$ symmetric $\varphi^4$ lattice model.
Theoretical studies (beyond mean-field theory) of the Casimir force in such systems  have been restricted to $\infty^{d-1}\times L$ (film) geometry within the $\varepsilon=4-d$ expansion above and at $T_c$ \cite{KrDi92a,DiGrSh06,GrDi07}. The most interesting region, however, is the region {\it below} $T_c$ where the scaling function of the Casimir force displays a characteristic minimum as detected by MC simulations \cite{dan-krech,vasilyev2009}. A basic difficulty in treating an infinite film system (for $n=1$) below $T_c$ is the existence of a film transition at a separate critical temperature
$T_{c,film} < T_c$ just in the region of the minimum of the Casimir force \cite{film}. A quantitative theory of the corresponding dimensional crossover between different critical behavior of the $d$ dimensional bulk transition at $T_c$ and the $d-1$ dimensional film transition at $T_{c,film}$ constitutes an as  yet unsolved problem, except for the case of the Gaussian model \cite{kastening-dohm}.

In the present paper we circumvent the problem of dimensional crossover by studying a  {\it finite} $L_\parallel^{d-1} \times L$ block geometry  with finite aspect ratio $\rho=L/L_\parallel$ \cite{dohm2009}. This geometry includes slab ($0<\rho < 1$), cubic ($\rho=1$), and  rod-like ($\rho > 1$) geometries (Fig. 1).
\begin{figure}[!ht]
\begin{center}
\subfigure{\includegraphics[clip,width=7.92cm]{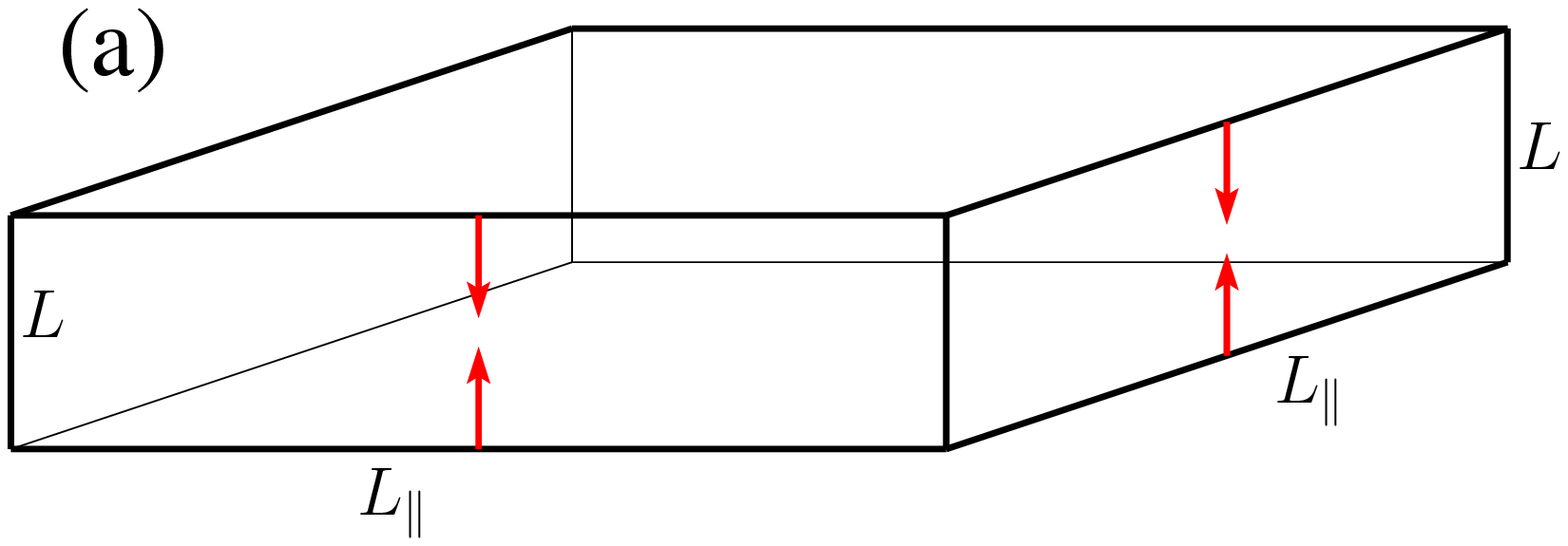}}
\subfigure{\includegraphics[clip,width=2.98cm]{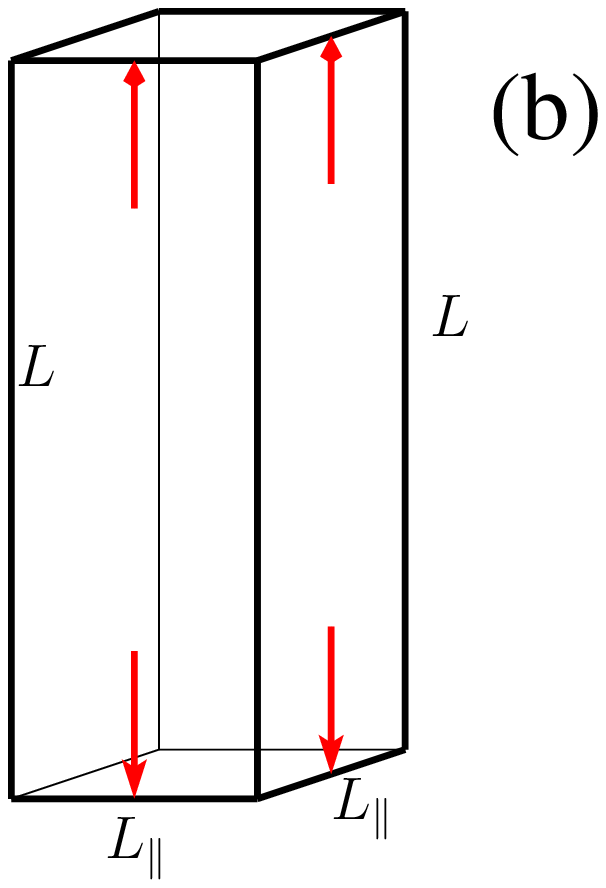}}
\end{center}
\caption{(Color online) Three-dimensional $L_\parallel^2 \times L$ block geometries with aspect ratio $\rho=L/L_\parallel$: (a) slab ($\rho < 1$), (b) rod ($\rho > 1$). Arrows: critical Casimir force (\ref{3d}). Our theory (in Secs. III - V) predicts that, for isotropic systems with periodic b.c. and short-range interactions, $F_{Casimir,s}$ at bulk $T_c$ is negative (attractive) for a slab, positive (repulsive) for a rod, and zero for a cube $(\rho=1)$.}
\end{figure}
The practical relevance of the slab geometry is based on the facts (i) that all experiments and all MC simulations have necessarily been performed at finite $\rho$ rather than at $\rho=0$, (ii) that the shape of the finite-size scaling function of the Casimir force depends on the aspect ratio $\rho$ only weakly in the
regime $\rho \ll 1$ \cite{dan-krech,vasilyev2009}, and (iii) that the singularity of the free energy and the Casimir force at $T_{c, film}$ for $\rho=0$ is only very weak
such that recent MC data \cite{vasilyev2009} could not detect this singularity. (By contrast, the logarithmic divergence of the specific heat for $\rho=0$ near $T_{c, film}$ should be well detectable.) This justifies
to describe the main features of the film system above {\it and below} $T_c$, to a good approximation, by a finite slab
geometry with small but finite aspect ratio. As an interesting by-product of our theory, the dependence on the aspect ratio for larger $\rho$ is obtained which permits us to describe the geometric crossover from slab ($\rho \ll 1$) over cubic ($\rho =1$) to rod-like ($\rho \gg 1$) geometries. The geometric crossover from block to cylindrical ($\rho = \infty$) geometries has been studied earlier near first-order transitions by Privman and Fisher \cite{Privman-Fisher}. In this context we note that systems with $\rho \gg 1$ are of experimental interest in the area of finite-size effects near the superfluid transition of $^4$He \cite{dohm,gasparini}. Furthermore, finite-size theories for block geometries are directly testable by MC simulations.

The finite-block system is  conceptually simpler
than the infinite film system because of a considerable technical advantage: For finite  $0 < \rho <\infty$, the system has a discrete mode spectrum with only one single lowest mode, in contrast to the more complicated situation of a lowest-mode {\it continuum} in film ($ \rho =0$) or cylinder ($ \rho =\infty$) geometry. This opens up the opportunity of building upon the  advances that have been achieved in the description of finite-size effects in systems that are finite in all directions \cite{dohm2008,BZ,RGJ,EDHW,EDC}. It is not clear {\it a priori}, however, in what range of $\rho$ such a theory is reliable since, ultimately, for sufficiently small  $\rho \ll 1$ or sufficiently large $\rho \gg 1$, the concept of separating a single lowest mode must break down. Therefore, as a crucial part of our theory, it is necessary to provide quantitative evidence for the expected range of applicability of our theory at finite $\rho$. This is one of the reasons why we also consider (in Sec. III) the large - $n$ limit whose exact results turn out to be helpful in estimating the range of validity of our approximate results for $n=1$.

As we shall present {\it three} different perturbation approaches with different ranges of applicability for the case $n=1$ we give here a brief overview of our strategy. The basic physical quantity is the singular part of the excess free energy density $f^{ex}_s$.
On the basis of previous work \cite{dohm2008,dohm2010} it is expected that, for our isotropic system with a finite volume $L_\parallel^{d-1} \times L$,  there exist three different regimes (a), (b), and (c) for the finite-size critical behavior of the excess free energy $f^{ex}_s$. These different regimes correspond to the three regions in Fig. 2 that are separated by the dashed lines.
\begin{figure}[!h]
\includegraphics[width=80mm]
{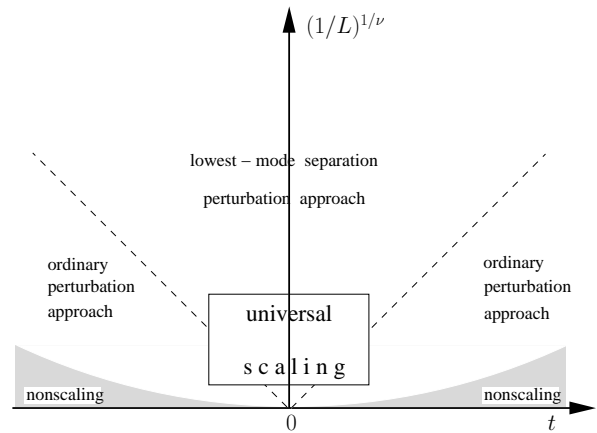} \caption{Schematic plot of the asymptotic part of the $L^{-1/\nu} -
t$ plane, with $t\equiv (T-T_c)/T_c$, for the $\varphi^4 $ model with  $n=1$ and with isotropic interactions on a simple-cubic lattice with lattice spacing $\tilde a$ in
a $L^{d-1}_\parallel \times L$ block geometry with periodic boundary conditions and finite aspect ratio $\rho = L/L_\parallel$. In the central
finite-size region (above the dashed lines), the lowest mode must
be separated whereas outside this region (below the dashed lines), ordinary perturbation
theory is applicable. Above the shaded region, universal finite-size scaling
is valid for isotropic systems. In the large - $L$ regime at $t\neq 0$ (shaded region), the exponential form of the size dependence violates both
finite-size scaling and universality because of a non-negligible dependence on the lattice spacing $\tilde a$. An analogous plot applies to the   $L_\parallel^{-1/\nu} -
t$ plane.  }
\end{figure}

(a) The regime well above $T_c$ that includes an exponential size
dependence $f^{ex}_s\sim \exp (- L /
\xi_{{\bf e}+})$  or $f^{ex}_s\sim \exp (- L_\parallel /
\xi_{{\bf e}+})$ for large
$L/\xi_{{\bf e}+} \gg 1$ and $L_\parallel/\xi_{{\bf e}+} \gg 1$  with
$\xi_{{\bf e}+}$ being the exponential ("true") bulk correlation length \cite{dohm2008,cd2000-2,fish-2}
above $T_c$; in this regime, $f^{ex}_s$ is expected to tend to {\it zero} in the high-temperature limit at finite $L$ and $L_\parallel$ or  in the limit of large $L$ and $L_\parallel$ at fixed temperature $T > T_c$.

(b) The central finite-size regime near $T_c$ that includes
the power-law behavior $f^{ex}_s \sim L^{-d}$ or $f^{ex}_s \sim L_\parallel^{-d}$ for large
$L$ at fixed $L/\xi_\pm \; , 0 \leq L/\xi_\pm \leq O (1)$ and for large
$L_\parallel$ at fixed $L_\parallel/\xi_\pm \; , 0 \leq L_\parallel/\xi_\pm \leq O (1)$ above, at
and below $T_c$ where $\xi_\pm$ is the second-moment bulk correlation
length.

(c) The regime well below $T_c$ that includes an exponential size dependence $\sim \exp (- L /
\xi_{{\bf e}-})$ or $\sim \exp (- L_\parallel /
\xi_{{\bf e}-})$ for large
$L/\xi_{{\bf e}-} \gg 1$ and $L_\parallel/\xi_{{\bf e}-} \gg 1$  with $\xi_{{\bf e}-}$ being the exponential
("true") bulk correlation length \cite{dohm2008,cd2000-2,fish-2} below $T_c$; in this regime, $f^{ex}_s$ is expected to have an exponential decay towards a {\it finite} value $-V^{-1} \ln 2$ in the low-temperature limit at finite volume $V$ \cite{dohm2010} and to tend to zero  in the limit of large volume at fixed
temperature $T < T_c$.

For a description of the
cases (a) and (c), ordinary perturbation theory with respect to the four-point coupling
$u_0$ of the $\varphi^4$ theory is sufficient. This ordinary perturbation approach is applicable to the regions below the dashed lines in Fig. 2. This approach will be presented in Sec. VI. For the
case (b), a separation of the lowest mode and a perturbation
treatment of the higher modes is necessary \cite{BZ, RGJ, EDHW,
EDC, dohm2008}. This approach is applicable to the region between the dashed lines in Fig. 2 which we refer to as the {\it central finite-size regime}. In Secs. IV and V we treat the
case (b) in $2<d<4$ dimensions on the basis of the  $\varphi^4$ lattice model in the minimal renormalization scheme at fixed dimension $d$ \cite{dohm1985}. We consider a simple-cubic lattice with isotropic short-range interactions. We shall demonstrate that our different perturbation approaches complement each other and that the corresponding results match reasonably well at intermediate values of the scaling variables. As will be shown in Secs. V and VI, these results agree well with Monte Carlo data for the three-dimensional Ising model by Hasenbusch for $\rho=1$ \cite{hasenbusch} and by Vasilyev et al.  for $\rho=1/6$ \cite{vasilyev2009} above, at, and below $T_c$.

We shall see that in all regimes (a)-(c), universal finite-size scaling \cite{pri} of isotropic systems is valid, except for the regions that are indicated by the shaded
areas in Fig. 2 where the exponential form of the size dependence violates both
finite-size scaling and universality because of a non-negligible dependence on the lattice spacing $\tilde a$. The boundaries of the shaded areas are not sharply defined; they are approximately determined by $L \simeq  24  \xi_\pm^3 / \tilde a^2$. This issue will be discussed in Sec. VI D.

\renewcommand{\thesection}{\Roman{section}}
\renewcommand{\theequation}{2.\arabic{equation}}
\setcounter{equation}{0}
\section*{II. Model and basic definitions}

We start from the O$(n)$ symmetric $\varphi^4$ lattice Hamiltonian
divided by $k_B T$
\begin{eqnarray}
\label{2a} H  &=&   \tilde a^d \Bigg[\sum_{i=1}^N
\left(\frac{r_0}{2} \varphi_i^2 + u_0 (\varphi_i^2)^2  \right) +
\sum_{i, j=1}^N \frac{K_{i,j}} {2} (\varphi_i - \varphi_j)^2
\Bigg], \;\nonumber\\
\end{eqnarray}
\be
r_0(T) = r_{0c} + a_0 t, \;\;t = (T - T_c) / T_c,
\ee
with $a_0>0$,
$u_0>0$ where $T_c$ is the {\it bulk} critical temperature. The
variables $\varphi_i \equiv \varphi ({\bf x}_i)$ are $n$-component
vectors which represent the internal degrees of freedom of $N$ particles on $N$ lattice points ${\bf x}_i$ of a $d$-dimensional
simple-cubic lattice with lattice constant $\tilde a$ and with periodic boundary conditions.
The components
$\varphi_i^{(\mu)} \; , \mu = 1, 2, \ldots, n$ of $\varphi_i$ vary
in the continuous range $- \infty \leq \varphi_i^{(\mu)} \leq
\infty$. We consider a finite
rectangular   $L_\parallel^{d-1} \times L$ block geometry of volume $V =  L_\parallel^{d-1} L $
with the aspect
ratio
\begin{equation}
\label{2aa}  \rho = \frac{L}{L_\parallel}\;.
\end{equation}
This block geometry includes the shape of a cube ($\rho=1$), of a finite slab ($0< \rho < 1$), and of a finite rod ($1 < \rho < \infty$) (Fig.1).

The free energy per component and per unit volume
divided by $k_BT$ is
\begin{equation}
\label{2b} f(t, L, L_\parallel) =  - (nV)^{-1} \ln Z (t, L,  L_\parallel)\;
\end{equation}
where
\begin{equation}
\label{2c} Z(t, L, L_\parallel) = \left[\prod_{i = 1}^N \frac{\int
d^n \varphi_i}{\tilde a^{n (2-d) / 2}} \right] \exp \left(- H \right)
\end{equation}
is the dimensionless partition function.
The bulk free
energy density per component divided by $k_B T$ is obtained by
\begin{equation}
\label{2cc}f_b(t) =\lim_{ L \to \infty} \lim_{L_\parallel  \to \infty}f(t, L, L_\parallel).
\end{equation}
The {\it film} free
energy density per component divided by $k_B T$ is obtained by taking the limit
$ L_\parallel \to \infty$ at fixed finite $L$ (i.e., $ \rho \to 0$)
\begin{equation}
\label{2i}f_{film}(t, L) = \lim_{ L_\parallel \to \infty}f(t, L, L_\parallel),
\end{equation}
corresponding to an  $\infty^{d-1} \times L$ geometry. Our model also includes  the limit
$ L \to \infty$ at fixed finite $L_\parallel$ (i.e., $ \rho \to \infty$) corresponding
to an  $L_\parallel^{d-1} \times \infty$ geometry which we shall refer to as  cylinder
geometry,
\begin{equation}
\label{2ii}f_{cyl}(t, L_\parallel) = \lim_{ L \to \infty}f(t, L, L_\parallel).
\end{equation}
The excess free energy density per component divided by $k_B T$  is
\begin{equation}
\label{2j}f^{ex}(t, L, L_\parallel) =  f(t, L,  L_\parallel) - f_b(t).
\end{equation}
For the finite $L_\parallel^{d-1} \times L$ system  we
define the Casimir force per unit area and per component in the $d^{th}$ (vertical) direction (Fig.1) as
\begin{equation}
\label{2k} F_{Casimir}(t, L,  L_\parallel) =  - \frac{\partial[L
f^{ex}(t, L,  L_\parallel) ]}{\partial L}
\end{equation}
where the derivative is taken at fixed $ L_\parallel$. There exist, of course, also Casimir forces  in the $d-1$ horizontal directions. Our approach is well suitable to calculate such forces. This will not be performed in this paper.

An important simplification of our model Hamiltonian (\ref{2a}) is the assumption of a rigid lattice with a rigid rectangular shape representing an idealized model system with a vanishing compressibility. The same assumption is made in lattice models on which previous Monte Carlo simulations of the Casimir force are based (see \cite{krech,toldin,vasilyev2009}). As a consequence, the number $N$ of particles and the length $L$ are directly related by $N=L_\parallel^{d-1}L /\tilde a^d$. Thus the derivative with respect to $L$ (at fixed $L_\parallel$, fixed $\tilde a$ and at fixed couplings $K_{ij}$ and $u_0$) in (\ref{2k}) is equivalent to a derivative with respect to the number of horizontal layers of the lattice, i. e., number of particles. Such a definition of the Casimir force is appropriate when the ordering degrees of freedom (particles in fluid films \cite{krech,toldin,vasilyev2009} or Cooper pairs in superconducting films \cite{wil-1}) can move in and out of the film system without significantly changing the mean interparticle distance in the system.  The definition (\ref{2k}) is, however, not appropriate for systems with a {\it fixed} number of ordering degrees of freedom. It appears that this is  reason why it was claimed in \cite{toldin} that the Casimir force "is not a measurable quantity for magnets". For similar claims see \cite{diehl}. There exist, however, (long-ranged) critical fluctuations of {\it elastic} degrees of freedom coupled to the order parameter in condensed matter systems with a finite compressibility (such as magnetic materials \cite{alpha}, alloys \cite{onukiBook}, and solids with structural phase transitions \cite{bruce-1}) which give rise to $L$ dependent critical Casimir forces that are not contained in a description based on rigid-lattice models of the type (\ref{2a}). A description of such thermodynamic Casimir forces is provided by coupling the variables $\varphi_i$ in (\ref{2a}) to the  elastic degrees of freedom  \cite{alpha,onukiBook,onuki}, in which case the free energy density $ f(t, L,  L_\parallel, N)$ depends on both the length $L$ and the number $N$ of particles as {\it independent} thermodynamic variables.  Such a model is relevant for the calculation of an $L$ dependent elastic response to the critical Casimir force (e.g., an $L$ dependent contribution to magnetostriction). In the case of a compressible system whose number $N$ of particles is fixed, the $L$ dependent thermodynamic Casimir force is given by
\begin{equation}
\label{2kkkk} F_{Casimir}(t, L,  L_\parallel,N) =  - \frac{\partial[L
f^{ex}(t, L,  L_\parallel,N) ]}{\partial L}
\end{equation}
where now the derivative is taken at fixed $t, L_\parallel$, and $N$. In such systems, anisotropy effects on the critical Casimir force are expected to play an important role. Eq. (\ref{2kkkk}) complements our arguments presented in \cite{real}. As we consider the thermodynamic Casimir effect as a {\it finite-size} effect, our definition  (\ref{2kkkk}) does not include the bulk part of the total thermodynamic force $- \partial[L
f(t, L,  L_\parallel,N) ]/\partial L$ which may give rise to measurable elastic bulk effects (such as a bulk contribution, e.g., to magnetostriction). Here we shall not further discuss this extension to systems with a finite compressibility which is beyond the scope of the present paper whose focus is on isotropic incompressible systems.

For small $|t|$, the bulk free energy density (\ref{2cc}) can be uniquely decomposed into singular and nonsingular parts
\begin{equation}
\label{3c}f_b(t) =  f_{b,s}(t) + f_{b,ns}(t).
\end{equation}
For large $L/\tilde a$, $L_\parallel/\tilde a$ and small $|t|$, a corresponding assumption is made \cite{priv} for
\begin{equation}
\label{3a} f(t, L,  L_\parallel) = f_{s}(t, L,  L_\parallel) + f_{ns}(t, L,  L_\parallel)
\end{equation}
and for the excess free energy
\begin{equation}
\label{3a} f^{ex}(t, L,  L_\parallel) = f^{ex}_{s}(t, L,  L_\parallel) + f^{ex}_{ns}(t, L,  L_\parallel)
\end{equation}

where $f_{ns}(t, L,  L_\parallel)$ and  $f^{ex}_{ns}(t, L,  L_\parallel)$ are regular functions of $t$  and where $f_{ns}(t, L, L_\parallel)$ remains regular in the bulk limit, $f_{ns}(t,L,L_\parallel) \to f_{b,ns}(t)$, whereas  $f_s(t,L,L_\parallel) \to f_{b,s}(t)$ becomes singular in this limit. It has been assumed \cite{pri} that, for periodic b.c.,  $f_{ns}(t, L,  L_\parallel)$ is independent of $L$ and $L_\parallel$, i. e., that it
is identical with the regular bulk part
$f_{b,ns}(t)$ of $f_b(t)$. We do not know of a general proof of this property; it appears to be valid for the $\varphi^4$ theory in the presence of short-range interactions $K_{i,j}$ but not in the presence of long-range correlations
\cite{dohm2008}.  The  critical
behavior of  $F_{Casimir}$ can be calculated from its singular part
\begin{equation}
\label{3d} F_{Casimir,s}(t, L, L_\parallel) =  - \frac{\partial[L
f^{ex}_{s}(t, L, L_\parallel) ]}{\partial L} .
\end{equation}
For the $\varphi^4$ lattice model with the interaction  given in (\ref{2g}) below it is expected that $f^{ex}_{ns}(t, L,  L_\parallel)=0$, thus $F_{Casimir}=F_{Casimir,s}$ which is consistent with our results in Secs. III - VI (see also the remark after Eq. (4.36) of \cite{dohm2008}).

Our main goal is to study the case $n=1$ at fixed finite aspect ratio $0< \rho < \infty$ including  extrapolations to the film and cylinder geometries. For comparison and as a guide for our extrapolations we shall also consider the exactly solvable limit $n \to \infty$.

For periodic boundary conditions, the Fourier representations are
\begin{equation}
\label{2d} \varphi({\bf x}_j) =
V^{-1} \sum_{\bf k} e^{i {\bf k} \cdot {\bf x}_j} \hat
\varphi({\bf k})
\end{equation}
and
\begin{equation}
\label{2e} K_{i,j}=K({\bf x}_i - {\bf x}_j) \; = \; N^{-1}
\sum_{\bf k} e^{i{\bf k} \cdot ({\bf x}_i - {\bf x}_j)} \widehat K
({\bf k}) \;,
\end{equation}
where the summations $\sum_{\bf k}$ run over $N$ discrete vectors
${\bf k} \equiv (k_1, k_2, \ldots, k_d)$ with Cartesian components
$k_\alpha = 2 \pi m_\alpha /  L_\parallel,  \alpha = 1,2, \cdots, d-1$, and $k_d = 2 \pi m_d / L, m_\beta
= 0, \pm 1, \pm 2, \cdots, $ in the range $ - \pi / \tilde a
\leq k_\beta < \pi / \tilde a$, $\beta= 1,...,d$.   In terms of the
Fourier components the Hamiltonian reads
\begin{eqnarray}
\label{2f} H  = V^{-1} \sum_{{\mathbf k}} \frac{1}{2} [r_0 + \delta
\widehat K (\mathbf k)] \hat \varphi({\mathbf k}) \hat
\varphi({-{\mathbf k}})) \nonumber\\ + \;u_0 V^{-3}
\sum_{{\mathbf{kp}}{\mathbf q}} [\hat \varphi({\mathbf k}) \hat
\varphi({{\mathbf p}})] [\hat \varphi({{\mathbf q}}) \hat
\varphi({-{\mathbf k}-{\mathbf p}-{\mathbf q}})] \qquad
\end{eqnarray}
where $ \delta \widehat K({\bf k}) = 2 [\widehat K({\bf 0}) -
\widehat K ({\bf k})]$. We assume a simple ferromagnetic
nearest-neighbor interaction
\begin{equation}
\label{2g} \delta \widehat K ({\bf k}) =  \frac{2}{ \tilde a^2} \sum_{\alpha =
1}^d  \left[1 - \cos (\tilde a  k_\alpha) \right] \;
\end{equation}
which has the isotropic long-wavelength form
\begin{equation}
\label{2h} \delta \widehat K ({\bf k})  \;= \;{\bf k}^2 \;\; +\;\; O
(k^4) \;.
\end{equation}
Thus it is appropriate to define a single second-moment bulk correlation length $\xi^\pm$ above (+) and below($-$) $T_c$ (see, e.g., Eq. (3.4) of \cite{dohm2008}). As a reference length that is finite for both $n=1$ and $n=\infty$ we shall use the asymptotic amplitude $\xi_{0+}$ of the second-moment bulk correlation length above $T_c$
\begin{equation}
\label{3f}\xi^+  = \xi_{0+} t^{-\nu}.
\end{equation}
For small $|t|$, the  asymptotic bulk power law is $f_{b,s}(t)=A^\pm |t|^{d\nu}$. Due to two-scale factor universality for isotropic systems \cite{dohm2008}, this can be written as
\begin{equation}
\label{3g}  f_{b,s}(t) = \left\{
\begin{array}{r@{\quad\quad}l}
                Q_1 \; (\xi_{0+}^{-1} t^{\nu})^d \;   & \mbox{for} \;\;\;  T > T_c\;, \\
 (A^-/A^+)Q_1 \; (\xi_{0+}^{-1} |t|^{\nu})^d \; &
\mbox{for} \;\;\;
                 T < T_c \;,
                \end{array} \right.
\end{equation}
with a universal constant $Q_1$ and the universal ratio of the
specific-heat amplitudes $A^-/A^+$.

The finite-size scaling form of the singular part of the free energy density
is, for isotropic systems in the asymptotic critical region $|t|\ll 1, L/\tilde a \gg 1, L_\parallel /\tilde a \gg 1$, \cite{dohm2008,pri}
\begin{equation}
\label{3h}f_{s}(t, L, L_\parallel)=L^{-d} F(\tilde x, \rho),
\end{equation}
where $ F(\tilde x, \rho)$ is a
dimensionless scaling function with the scaling variable
\begin{equation}
\label{3j}\tilde{x}=t(L/\xi_{0+})^{1/\nu}.
\end{equation}
The bulk part $F^\pm_b(\tilde x)$ of  $F(\tilde{x},\rho)$  is obtained from (\ref{3g}) and (\ref{3h}) in the limit of large $|\tilde x|$ as $L^d f_{s}(t, L, L_\parallel) \to F^\pm_b(\tilde{x})$ with
\begin{eqnarray}
\label{3jj}
F^\pm_b(\tilde{x}) =\left\{
\begin{array}{r@{\quad \quad}l}
                         \; Q_1 \tilde x^{d \nu}\quad          & \mbox{for} \;T > T_c\;, \\
                         \; (A^-/A^+)Q_1\mid\tilde{x}\mid^{d\nu}& \mbox{for} \;T <
                 T_c .\;
                \end{array} \right.
\end{eqnarray}
This implies the scaling form
\begin{eqnarray}
\label{3k}
f^{ex}_{s}(t, L,  L_\parallel)= L^{-d}F^{ex}(\tilde{x},\rho),
\end{eqnarray}
\begin{eqnarray}
\label{3kk}
F^{ex}(\tilde{x},\rho)= F(\tilde{x},\rho) - F^\pm_b(\tilde{x}).
\end{eqnarray}
Together with  (\ref{3d}), this leads to the scaling form of  the critical Casimir force for systems with isotropic interactions
\begin{equation}
\label{3l} {F}_{Casimir,s}(t, L,  L_\parallel)=L^{-d}X(\tilde x, \rho)
\end{equation}
with the scaling function
\begin{equation}
\label{3nn} X(\tilde{x},\rho) =(d-1)  F^{ex} (\tilde{x},\rho) -
\frac{\tilde{x}}{\nu}
\frac{\partial F^{ex}(\tilde{x},\rho)}{\partial\tilde{x}}- \rho
\frac{\partial F^{ex}(\tilde{x},\rho)}{\partial \rho}.
\end{equation}
These scaling functions have finite limits for  $\rho \to 0$ at fixed $L$ and  at fixed $\tilde{x}$ corresponding to film geometry,
\begin{equation}
\label{3no}f_{film,s}(t, L)=L^{-d} F_{film}(\tilde x),
\end{equation}
\begin{eqnarray}
\label{3qqq}
F_{film}(\tilde{x})= \lim_{\rho \to 0}F(\tilde{x},\rho),
\end{eqnarray}
\begin{eqnarray}
\label{3qq}
F^{ex}_{film}(\tilde{x})= \lim_{\rho \to 0}F^{ex}(\tilde{x},\rho),
\end{eqnarray}
\begin{eqnarray}
\label{3Xqq}
X_{film}(\tilde{x})= \lim_{\rho \to 0}X(\tilde{x},\rho) \nonumber \\=(d-1)  F_{film}^{ex} (\tilde{x}) -
\frac{\tilde{x}}{\nu}
\frac{\partial F_{film}^{ex}(\tilde{x})}{\partial\tilde{x}}.
\end{eqnarray}
Note that $\nu$ denotes the {\it bulk} critical exponent and $\tilde{x}$ is the scaling variable with respect to  {\it bulk} criticality, not with respect to film criticality.

In a rod-shaped geometry with $\rho \gg 1$, a representation of the scaling form in terms of the length $L_\parallel$ and of the scaling variable
\begin{equation}
\label{3o}\tilde{x}_\parallel=t(L_\parallel/\xi_{0+})^{1/\nu},
\end{equation}
rather than in terms of $\tilde x$, is more appropriate. Because of
\begin{equation}
\label{3p}
\tilde{x} = \tilde{x}_\parallel \rho^{1/\nu}
\end{equation}
we obtain from (\ref{3d}), (\ref{3h}), (\ref{3j}), (\ref{3k}), and (\ref{3kk})
\begin{equation}
\label{3q}f_{s}(t, L, L_\parallel)=L_{\parallel}^{-d} \Phi(\tilde x_\parallel, \rho),
\end{equation}
\begin{equation}
\label{3qq}f^{ex}_{s}(t, L, L_\parallel)=L_{\parallel}^{-d} \Phi^{ex}(\tilde x_\parallel, \rho),
\end{equation}
\begin{equation}
\label{3qqq} {F}_{Casimir,s}(t, L,  L_\parallel)=L_\parallel^{-d}\Xi(\tilde x_\parallel, \rho),
\end{equation}
with the scaling functions
\begin{equation}
\label{3r}\Phi(\tilde x_\parallel, \rho) = \rho^{-d} F(\tilde x_\parallel \rho^{1/\nu}, \rho),
\end{equation}
\begin{equation}
\label{3rr}\Phi^{ex}(\tilde x_\parallel, \rho) = \rho^{-d} F^{ex}(\tilde x_\parallel \rho^{1/\nu}, \rho),
\end{equation}
\begin{eqnarray}
\label{3rrr}\Xi(\tilde x_\parallel, \rho)&=& \rho^{-d} X(\tilde x_\parallel \rho^{1/\nu}, \rho)
\nonumber \\ &=& -\Phi^{ex}(\tilde x_\parallel, \rho) \;+ \;(1/\rho)\;
\frac{{\partial \Phi^{ex}(\tilde x_{\parallel},\rho)}}{\partial (1/\rho)}.\;\;
\end{eqnarray}
It turns out that they have finite limits for $\rho \to \infty$ at fixed $L_\parallel$ and fixed $\tilde x_{\parallel}$ corresponding to cylinder geometry,
\begin{equation}
\label{3rrrr}f_{cyl,s}(t, L_{\parallel})=L_{\parallel}^{-d} \Phi_{cyl}(\tilde x_{\parallel}),
\end{equation}
\begin{equation}
\label{3s}\Phi_{cyl}(\tilde x_\parallel) =  \lim_{\rho \to \infty}\Phi(\tilde x_\parallel, \rho),
\end{equation}
\begin{equation}
\label{3ss}\Phi^{ex}_{cyl}(\tilde x_\parallel) =  \lim_{\rho \to \infty}\Phi^{ex}(\tilde x_\parallel, \rho),
\end{equation}
\begin{equation}
\label{3sss}\Xi_{cyl}(\tilde x_\parallel) =  \lim_{\rho \to \infty}\Xi(\tilde x_\parallel, \rho)= - \Phi^{ex}_{cyl}(\tilde x_\parallel).
\end{equation}
Quantitative results for the various scaling functions will be presented in Sec. III for $n=\infty$ and in Secs. V  and VI for $n=1$.

We recall that all finite-size scaling forms given in this and the subsequent sections are not valid in a small part of the asymptotic region for large $L$ (or large $L_{\parallel}$) at fixed  $t\neq 0$ in  the $L^{1/\nu} - t$ plane (or $L_{\parallel}^{1/\nu} - t$ plane) above and below $T_c$ (corresponding to the shaded regions in Fig. 2) where exponential nonscaling terms exist that depend explicitly on the lattice constant $\tilde a$. This issue will be discussed in Sec. VI.

\section*{III. Large - ${\bf n}$  limit }
\renewcommand{\thesection}{\Roman{section}}
\renewcommand{\theequation}{3.\arabic{equation}}
\setcounter{equation}{0}

\subsection*{A. Free energy density}
Generalizing Eq. (6.30) of Ref. \cite{dohm2008} to $L_\parallel^{d-1}\times L$ geometry we obtain for the free energy density per component in the limit $n \to \infty$ at fixed $u_0 n$
\begin{eqnarray}
\label{3a} &&f(t,L, L_\parallel) = \lim_{n \to \infty}[-(n
V)^{-1} \ln Z(t,L, L_\parallel)]  = - \frac{\ln (2
\pi)}{ 2 \tilde a^d} \nonumber\\&-& \frac{(r_0-\chi^{-1})^2}{16 u_0 n} +
\frac{1}{2V} {\sum_{\bf k}} \ln [(\chi^{-1} +  \delta
\widehat K (\mathbf k)])
\tilde a^2]
\end{eqnarray}
where $Z(t,L, L_\parallel)$ is defined by (\ref{2c})
and $\chi(t,L, L_\parallel)^{-1}$ is determined implicitly by
\begin{eqnarray}
\label{4b}
 \chi^{-1}= r_0 +  \frac{4 u_0n}{
V} {\sum_{\bf k}} [\chi^{-1} +
\delta
\widehat K (\mathbf k)]^{-1}.
\end{eqnarray}
The condition $\chi^{-1} = 0$ for bulk ($L \to \infty, L_\parallel \to \infty $) criticality yields the critical value of $r_0$ as
\begin{equation}
\label{3.3} r_{0c} = - 4 u_0 n \int_{\bf k} \delta
\widehat K (\mathbf k)^{-1}, \;\;
\int\limits_{\bf k} \; \equiv \; \prod^d_{\alpha = 1} \;
\int\limits^{\pi / \tilde a}_{- \pi / \tilde a} \; \frac{d
k_\alpha}{2\pi}.
\end{equation}
Eqs. (\ref{3a}) and (\ref{4b}) are valid for $T\geq T_c$ $(r_0 \geq r_{0c})$ and $T < T_c$  $(r_0 < r_{0c})$. For $T\geq T_c$, the quantity $\chi(t,L, L_\parallel)$ represents the susceptibility per component.

\subsection*{B. Exact scaling functions}
In the following we present  exact results for the finite-size scaling
functions $F$,  $\Phi$, $F^{ex}$,  $\Phi^{ex}$, $X$, and $\Xi$ in
$2 < d < 4$ dimensions. We rewrite (\ref{3a}) and (\ref{4b}) in terms of
$ r_0 - r_{0c} = a_0 t$ and assume large $L/\tilde a$, large $L_\parallel/\tilde a$, small $|r_0 - r_{0c}|/\tilde a^2 \ll 1$ and $|r_0 - r_{0c}|L^2 \lesssim O(1)$, $|r_0 - r_{0c}|L_\parallel^2  \lesssim O(1)$. Evaluation of the sums in (\ref{3a}) and (\ref{4b}) (see App. A) leads to the scaling form of $f_{s}(t,L,L_{\parallel})$, (\ref{3h}) and  (\ref{3j}),
where $\nu = (d-2)^{-1}$  and
\be
\xi_{0+} = \Big(\frac {4 u_0 n A_d}{\varepsilon a_0
}\Big)^{1/(d-2)}
\ee
with the geometric factor
\be
\label{4e} A_d \;=\; \frac{\Gamma(3-d/2)}{2^{d-2} \pi^{d/2}
(d-2)}.
\ee
For an arbitrary finite shape factor $0 < \rho  < \infty$ we find
the finite-size scaling function
\be
\label{3.9x}
 F (\tilde x, \rho) \; = \;  \frac{A_d}{2 \varepsilon}
\left[\tilde x P(\tilde x, \rho)^2 \; - \; \frac{2}{d} \; P(\tilde x, \rho)^d \right] + \frac{1}{2}{\cal G}_0( P(\tilde x, \rho)^2, \rho),
\ee
\begin{eqnarray}
\label{3.10x} {\cal G}_j( P^2, \rho)&=&  (4\pi^2)^{-j}\int\limits_0^\infty dz z^{j-1}
  \exp {\left(-\frac{P^2z}{4\pi^2}\right)}
  \nonumber\\ &\times& \left\{(\pi / z)^{d/2}-\Big[\rho K(\rho^2z)\Big]^{d-1}K(z)
    \right\} ,
\end{eqnarray}
where $P(\tilde x, \rho)$ is determined implicitly by
\be
\label{3.11x}
P^{d-2} \; = \; \tilde x  -   \frac{\varepsilon}{A_d}{\cal G}_1( P^2, \rho)
\ee
with
\begin{equation}
\label{funktionK} K(z) = \sum^{\infty}_{m= -\infty} \;\exp (- z
 m^2).
\end{equation}
The earlier result of Eqs. (6.32)-(6.34) of \cite{dohm2008} for cubic geometry is obtained from  (\ref{3.9x}) - (\ref{3.11x}) by setting $\rho=1$.
In the film limit $\rho \to 0$ at finite $L$,   we obtain $\rho K(\rho^2z)\to (\pi/z)^{1/2}$,
\begin{equation}
\label{3.10}
 F_{film} (\tilde x) \; = \;  \; \frac{A_d}{2 \varepsilon}
\left[\tilde x P(\tilde x)^2 \; - \; \frac{2}{d} \; P(\tilde x)^d
\right]  + \frac{1}{2}\;{\cal G}_{0,film}( P(\tilde x)^2),
\end{equation}
\begin{eqnarray}
\label{3.11} {\cal G}_{j,film}( P^2)&=& (4\pi^2)^{-j}\int\limits_0^\infty dz z^{j-1}
\exp {\left(-\frac{P^2z}{4\pi^2}\right)}
\Big(\frac{\pi}{z}\Big)^{(d-1)/2} \nonumber \\ &\times& \left\{(\pi / z)^{1/2}-K(z)
\right\} ,
\end{eqnarray}
where $P(\tilde x)$ is determined implicitly by
\begin{equation}
\label{3.12}
P^{d-2} \; = \; \tilde x \; - \;  \frac{\varepsilon}{A_d}\;{\cal G}_{1,film}( P^2) . \;\;
\end{equation}
For $n \to \infty$, a finite film-critical temperature $0<T_{c,film}(L)<
T_c$ exists only in $d>3$ dimensions whereas
$T_{c,film}(L)=0$ in $d\leq 3$ dimensions. Eqs. (\ref{3.10}) - (\ref{3.12}) are valid in the asymptotic region near bulk $T_c$.
\begin{figure}[!ht]
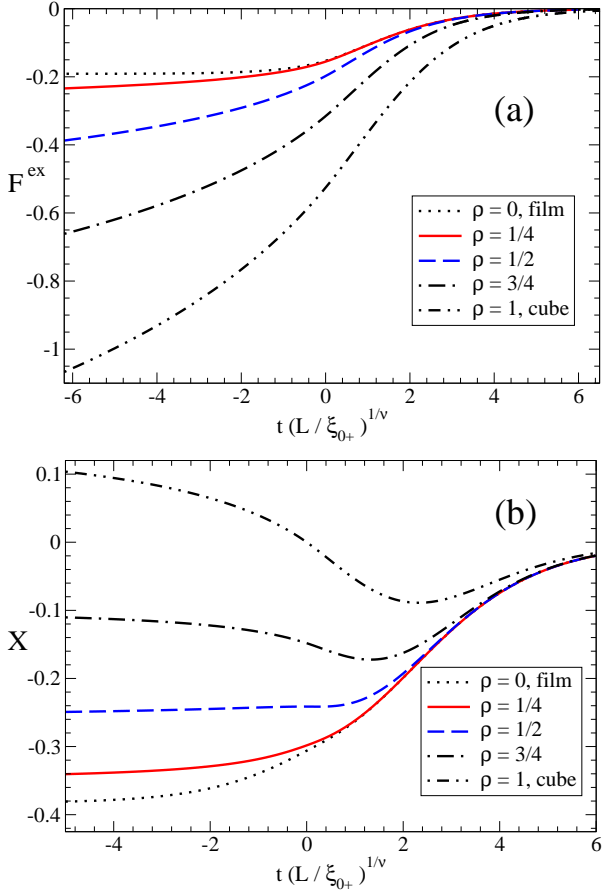

\begin{center}
\subfigure{\includegraphics[clip,width=7.92cm]{ExcessFreieEnergieSphKastenFilmWuerfelNeu.eps}}
\subfigure{\includegraphics[clip,width=7.92cm]{CasimirSphKastenFilmWuer.eps}}
\end{center}
\caption{(Color online) Scaling functions (a) $F^{ex}(\tilde x, \rho)$, (\ref{3kk}), (\ref{3.9x}) - (\ref{3.11x}), and (b) $X(\tilde x, \rho)$, (\ref{3nn}), as a function of $\tilde x = t (L/\xi_{0+})^{1/\nu}$ in the
large-$n$ limit in three dimensions for film geometry ($\rho=0$,  dotted lines),
for slab geometry  (solid lines: $\rho= 1/4$ , dashed lines: $\rho= 1/2$, dot-dashed lines: $\rho= 3/4$), and for cubic geometry  ($\rho=1$, double-dot-dashed lines). For $\tilde x \to - \infty$, the curves in (a) with $\rho>0$ diverge logarithmically towards $-\infty$ whereas the $\rho=0$ curve in (a) has a finite low-temperature limit -0.191.  All curves in (b) have finite limits for $\tilde x\to -\infty$ as given by (\ref{4i}) and (\ref{4ii}).}
\end{figure}
\begin{figure}[!ht]
\begin{center}
\subfigure{\includegraphics[clip,width=7.92cm]{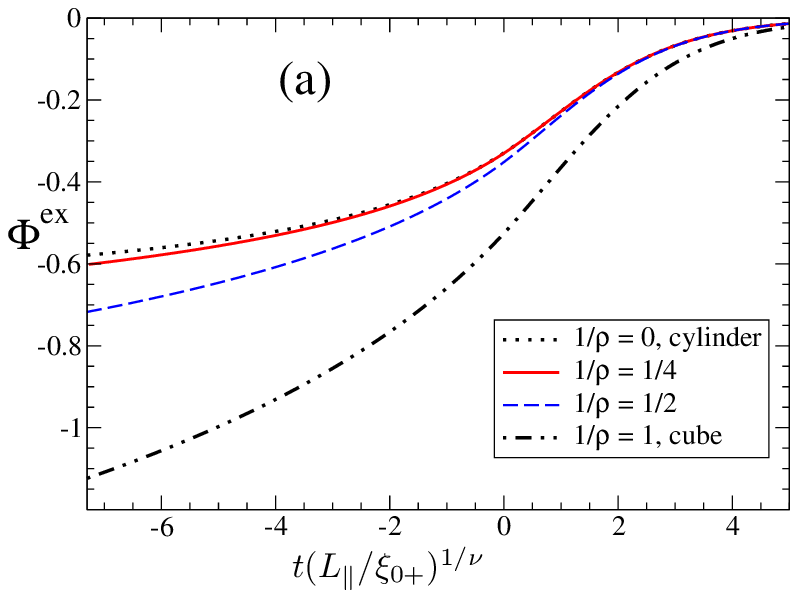}}
\subfigure{\includegraphics[clip,width=7.92cm]{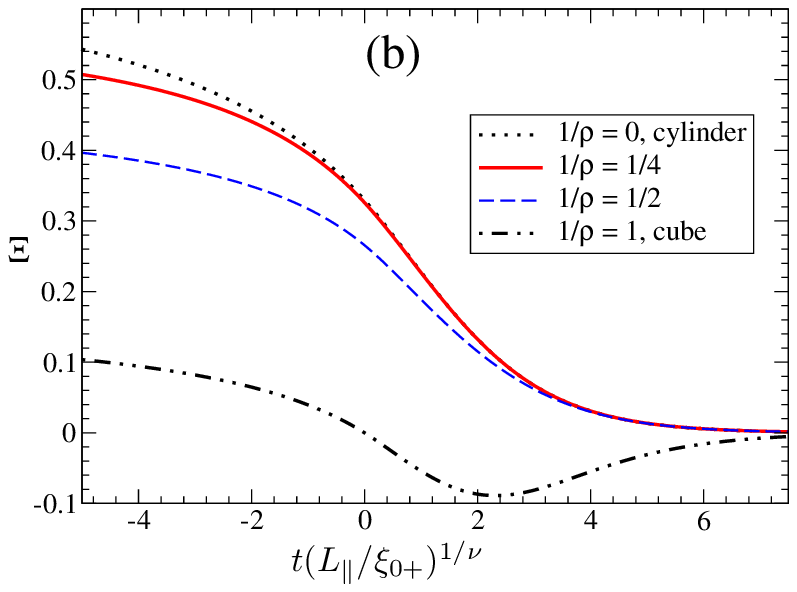}}
\end{center}
\caption{(Color online)  Scaling functions  (a) $\Phi^{ex}(\tilde x_\parallel, \rho)$, (\ref{3rr}),   and (b) $\Xi(\tilde x_{\parallel}, \rho)$, (\ref{3rrr}),  as a function of $\tilde{x}_\parallel=t(L_\parallel/\xi_{0+})^{1/\nu}$ in the
large-$n$ limit in three dimensions for cylinder  geometry ($1/\rho=0$, dotted lines), for rod geometry (solid lines: $1/\rho = 1/4$,  dashed lines: $1/\rho =1/2$), and for cubic geometry  ($1/\rho=1$, double-dot-dashed lines). For $\tilde x_\parallel \to - \infty$, the curves in (a) with $1/\rho>0$ diverge logarithmically towards $-\infty$ whereas the $1/\rho=0$ curve in (a) has a finite low-temperature limit -0.719.   All curves in (b) have finite limits for $\tilde x_{\parallel}\to -\infty$ as given by (\ref{4iii}) and (\ref{4iiii}).}
\end{figure}

In the cylinder limit $\rho \to \infty$ at finite $L_{\parallel}$,   we obtain $ K(\rho^2z)\to 1$, $P(\tilde x, \rho)/\rho \to P_{cyl}(\tilde x_{\parallel})$,
\begin{eqnarray}
\label{scal.cyl}
\Phi_{cyl}(\tilde x_\parallel) &=& \;  \frac{A_d}{2 \varepsilon}
\left[\tilde
x_{\parallel} P_{cyl}(\tilde
x_{\parallel})^2 \; - \; \frac{2}{d} \; P_{cyl}(\tilde
x_{\parallel})^d \right] \nonumber\\&+& \frac{1}{2}{\cal G}_{0,cyl}( P_{cyl}(\tilde x_\parallel)^2),
\end{eqnarray}
\begin{eqnarray}
\label{3.14} {\cal G}_{j,cyl}( P_{cyl}^2)&=& (4\pi^2)^{-j}\int\limits_0^\infty dz z^{j-1}
  \exp {\left(-\frac{z P_{cyl}^2}{4\pi^2}\right)}
  (\pi / z)^{1/2}\nonumber \\ &\times& \left\{(\pi / z)^{(d-1)/2}-[K(z)]^{d-1}
    \right\} ,
\end{eqnarray}
where $P_{cyl}(\tilde x_{\parallel})$ is determined implicitly by
\begin{equation}
\label{3.15}
P_{cyl}^{d-2} \; = \; \tilde
x_{\parallel}  - \;  \frac{\varepsilon}{A_d}\;{\cal G}_{1,cyl}( P_{cyl}^2) . \;\;
\end{equation}

In this cylinder limit, the system has an infinite extension only in the $d^{th}$  direction, i.e., it is essentially one-dimensional, thus no finite critical temperature exists in the cylinder case at finite $L_\parallel$. Eqs. (\ref{scal.cyl}) - (\ref{3.15}) are valid in the asymptotic region near  bulk $T_c$.

The bulk part $F_b^\pm(\tilde x)$ of $ F(\tilde x,\rho)$  in the large - $n$ limit is given by (\ref{3jj}) with $Q_1= (d - 2) A_d/[2 d (4 - d)]$ and $A^-/A^+=0$. Correspondingly, the bulk part of $ \Phi(\tilde x_\parallel,\rho)$  is $\Phi_b^\pm(\tilde x_\parallel)=F_b^\pm(\tilde x_\parallel)$. From (\ref{3kk}) and (\ref{3rr}) we then obtain the scaling functions  $F^{ex}$ and  $\Phi^{ex}$ of the excess free energy density.

The scaling functions $X$ and $\Xi$ of the Casimir force are obtained from $F^{ex}$ and  $\Phi^{ex}$  according to (\ref{3nn}), (\ref{3Xqq}), (\ref{3rrr}), and (\ref{3sss}). All scaling functions  are shown in Figs. 3 and 4 for several values of $\rho$ in three dimensions, illustrating the crossover from film geometry ($\rho = 0$, dotted lines in Fig. 3) over cubic geometry (double-dot-dashed lines) to cylinder geometry ($1/\rho = 0$, dotted lines in Fig. 4). We see that, for $O(-10)< \tilde x \leq \infty$ and $O(-10)< \tilde x_\parallel \leq \infty$, the scaling functions
for slab ($\rho < 1$) and rod ($1/\rho < 1$) geometries, respectively, provide a reasonable approximation for the
scaling functions  (i) for film geometry if
the shape factor $\rho$ is sufficiently small, (ii) for cylinder geometry if
the inverse shape factor $1/\rho$ is sufficiently small. This is not true, however, in the low-temperature limit $\tilde x \to -\infty$ and $\tilde x_\parallel \to -\infty$, respectively (see the following subsection).

\subsection*{C. Monotonicity properties at fixed ${\bf \rho}$}

For fixed $\rho$, $F^{ex}(\tilde x, \rho)$ and $\Phi^{ex} (\tilde x_\parallel, \rho)$ are monotonically increasing functions of $\tilde x$ and $\tilde x_\parallel$, respectively [see Figs. 3(a) and 4(a)]. They vanish exponentially fast for $\tilde x \to \infty$ and $\tilde x_\parallel \to \infty$  and have  logarithmic divergencies  towards $-\infty$ for $\tilde x \to -\infty$ and $\tilde x_\parallel \to -\infty$, respectively, for finite $0 < \rho < \infty$.

To derive the latter property, consider the quantity $P$ as determined by (\ref{3.11x}). It is finite and positive for $-\infty < \tilde x <\infty$ and vanishes for  $\tilde x \to -\infty$. More specifically, the function ${\cal G}_1$ has the divergent small-$P^2$ behavior ${\cal G}_1( P^2, \rho) \approx - \rho^{d-1}P^{-2}$ [see (\ref{Geinsasym}) in App. A]. According to (\ref{3.11x}), this implies  that $P^2$ vanishes as $P^2 \approx \varepsilon A_d^{-1}\rho^{d-1}(-\tilde x)^{-1}$ for $\tilde x \to -\infty$. Thus the behavior of $F(\tilde x, \rho)$, (\ref{3.9x}), for large negative $\tilde x$ is  given by
\begin{equation}
\label{4hhx}
F(\tilde x, \rho)=F^{ex}(\tilde{x},\rho)\approx  - \frac{1}{2}\rho^{d-1}+ \frac{1}{2}{\cal G}_0( P(\tilde x, \rho)^2, \rho)
\end{equation}
for  $-\tilde{x} \gg 1$. The function ${\cal G}_0$ has a divergent small-$P^2$ behavior  as given by (\ref{Gnullasym}) in App. A. The resulting logarithmic divergency is
\begin{equation}
\label{4hh}
F^{ex}(\tilde{x},\rho)\approx - \frac{1}{2} \rho^{d-1} \ln \Big(\frac{4\pi^2 A_d |\tilde x|}{\varepsilon \rho^{d-1}}\Big)  - \frac{1}{2}\rho^{d-1} + \frac{1}{2}{\cal C}_0(\rho)
\end{equation}
with ${\cal C}_0(\rho)$ given by (\ref{F.15xx}). For finite $0 < \rho < \infty$, Eqs. (\ref{3p}) and (\ref{3rr})  imply a corresponding logarithmic divergency of $\Phi^{ex}(\tilde{x_\parallel},\rho)$ for $\tilde{x_\parallel} \to -\infty$.

By contrast, we shall find a {\it nonmonotonic} dependence of the scaling functions  $F^{ex}$ and $\Phi^{ex}$ on $\tilde x$ and $\tilde x_\parallel$ for the $n=1 $ universality class for finite $0 < \rho < \infty$ in the central finite-size scaling regime described in Sec.  V   below.

For the film system in the large-$n$ limit, we confine ourselves to the case $ d = 3 $. We find from (\ref{3.11}) that ${\cal G}_{1,film}( P^2) \approx (4\pi)^{-1} \ln P^{2}$ for small $P^2$ and that $P^2$ vanishes as $P^2 \propto e^{\tilde x}$ for $\tilde x \to -\infty$. This implies that $F^{ex}_{film}(\tilde{x})$ has  a {\it finite} value in the low-temperature limit \cite{dan} [compare Fig. 3(a)]
\begin{equation}
\label{4hhh}
\lim_{\tilde{x} \to -\infty}F^{ex}_{film}(\tilde{x})= \frac{1}{2}\;{\cal G}_{0,film}(0)= -0.191 \;\; {\rm for} \;\; d=3.
\end{equation}

In the cylinder system in the large-$n$ limit, we find from (\ref{3.14}) that ${\cal G}_{1,cyl}( P_{cyl}^2) \approx -c_1 P_{cyl}^{-1}$ for small $P_{cyl}$ with $c_1=(4\pi)^{-1/2}\int_0^\infty dyy^{-1/2}e^{-y}>0$  and $P_{cyl} \approx c_1\varepsilon A_d^{-1} ({-\tilde x_\parallel})^{-1}$ for $\tilde x_\parallel \to -\infty$. This implies that, for $2<d<4$, $\Phi^{ex}_{cyl}(\tilde{x_\parallel})$ has  a {\it finite} value in the low-temperature limit
\begin{equation}
\label{4hhhhh}
\lim_{\tilde{x_\parallel} \to -\infty}\Phi^{ex}_{cyl} (\tilde x_\parallel) = \frac{1}{2}\;{\cal G}_{0,cyl}(0)
\end{equation}
with $\frac{1}{2}\;{\cal G}_{0,cyl}(0)= -0.719$ for $d=3$ [compare Fig. 4(a)].

In contrast to $F^{ex}$ and  $\Phi^{ex}$, the scaling functions $X$ and $\Xi$ turn out to be {\it nonmonotonic} functions of their scaling variables $\tilde x$ and $ \tilde x_\parallel$, respectively, in an intermediate range of $\rho$ where $\rho \sim O(1)$ [see Figs. 3(b) and 4(b)]. In this range, $X$ exhibits a change of sign near $\tilde x=0$: The Casimir force changes from an repulsive force below $T_c$ to an attractive force above $T_c$. Especially for $\rho=1, d=3$, this change of sign occurs exactly at $T_c$ where $X(0,1)=0$, $\Xi(0,1)=0$. In the range $\rho \lesssim 1/2$, $X<0$ is a monotonically increasing function of $\tilde x$. In the range $\rho \gtrsim 3/2$, $\Xi>0$ is a monotonically decreasing function of $\tilde x_{\parallel}$.

Above $T_c$,  $X$ and $\Xi$  have an exponential decay towards zero as functions of $\tilde x \gg 1$ and $ \tilde x_\parallel \gg 1$, respectively, as follows from the exponential decay of $F^{ex}$ and  $\Phi^{ex}$. Below $T_c$, the scaling functions $X$ and $\Xi$  have {\it finite} values in the low-temperature limits $\tilde x \to -\infty$ and $\tilde x_\parallel \to -\infty$, respectively, for all  $-\infty \leq \rho \leq \infty$, unlike the divergent behavior of $F^{ex}$ and  $\Phi^{ex}$ for finite $\rho$. To derive the low-temperature limit of $X$ we use (\ref{3kk}) and (\ref{3.9x}) to rewrite (\ref{3nn})  as
\begin{eqnarray}
\label{X-large-n} X(\tilde{x},\rho) =   F_b^\pm(\tilde{x}) + \frac{A_d}{\varepsilon}\Big[\frac{1}{2}\tilde x P^2- \frac{d-1}{d}P^d\Big] \nonumber\\+ \frac{1}{2}\Big[(d-1){\cal G}_0(P^2,\rho) -\rho \frac{\partial {\cal G}_0(P^2,\rho)}{\partial \rho}\Big] \;\;
\end{eqnarray}
with $P(\tilde{x},\rho)^2$ determined by (\ref{3.11x}). For $P^2 \to 0$, the divergent parts of the last two terms cancel each other which leads to a finite limit
\begin{eqnarray}
\label{4i}
\lim_{\tilde{x} \to -\infty}X(\tilde{x},\rho) = -\frac{1}{2}\rho^{d-1} &+& \frac{1}{2}\lim_{P \to 0} \Big[(d-1){\cal G}_0(P^2,\rho) \nonumber\\&-&\rho \frac{\partial {\cal G}_0(P^2,\rho)}{\partial \rho}\Big]
\end{eqnarray}
with a nontrivial $\rho$ dependence. Similarly we obtain a finite limit
\begin{equation}
\label{4iii}
\lim_{\tilde{x_\parallel} \to -\infty}\Xi (\tilde x_\parallel, \rho) = \rho^{-d}\lim_{\tilde{x} \to -\infty}X(\tilde{x},\rho).
\end{equation}
This is in contrast to the exponential decay of $X$ and $\Xi$ towards {\it zero} for $\tilde x \to -\infty$ and $\tilde x_\parallel \to -\infty$, respectively,  for the $n=1$ universality class that we shall find in Sec. VI below.
\begin{figure}[!ht]
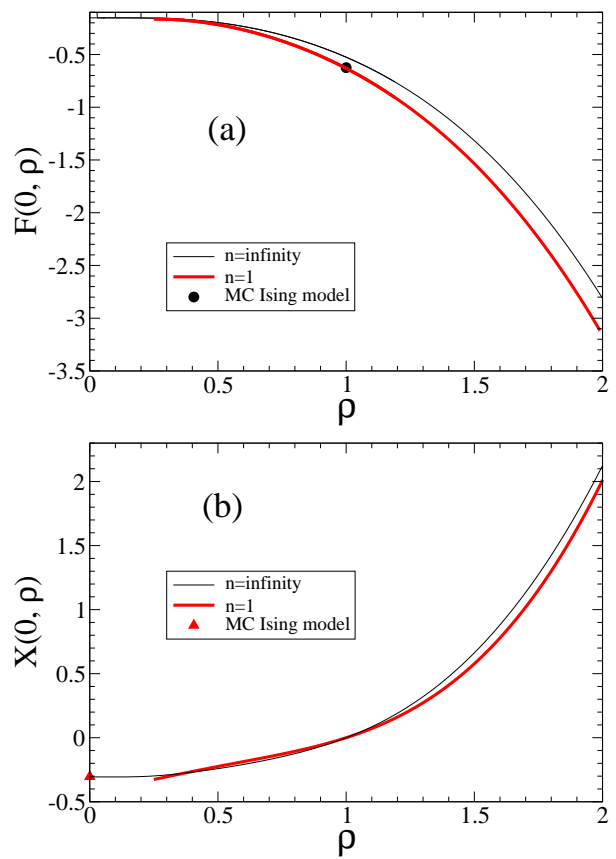

\begin{center}
\subfigure{\includegraphics[clip,width=7.92cm]{FreieEnergieSphIsingBeiTcOhnePunkte.eps}}
\subfigure{\includegraphics[clip,width=7.92cm]{CasimirSphIsingBeiTcKastenFilmBisRho2.eps}}
\end{center}
\caption{(Color online)  Critical amplitudes (a)  $F(0, \rho)$, (\ref{3kk}),  and (b) $X(0, \rho)$, (\ref{3nAmplitude}), at $T_c$ in three dimensions as a function of the aspect ratio $\rho$ in the
large-$n$ limit
[thin lines, from (\ref{3.9x})] and for $n=1$ [thick lines, from (\ref{6-2})]).
For $n=\infty$, $F_{film}(0)\equiv F(0,0)= -0.153$ and $X_{film}(0)\equiv X(0,0)= 2F_{film}(0)= -0.306$. At $\rho=1$, $X(0, 1)$ vanishes for both $n=\infty$ and $n=1$.
Monte Carlo data  of the $d=3$ Ising model for $\rho=1$ by Mon \cite{mon-1} [full circle in (a)] and for $\rho=0$
by Vasilyev et al. \cite{vasilyev2009} [triangle in (b)]. See also Fig. 7.}
\end{figure}
\begin{figure}[!ht]
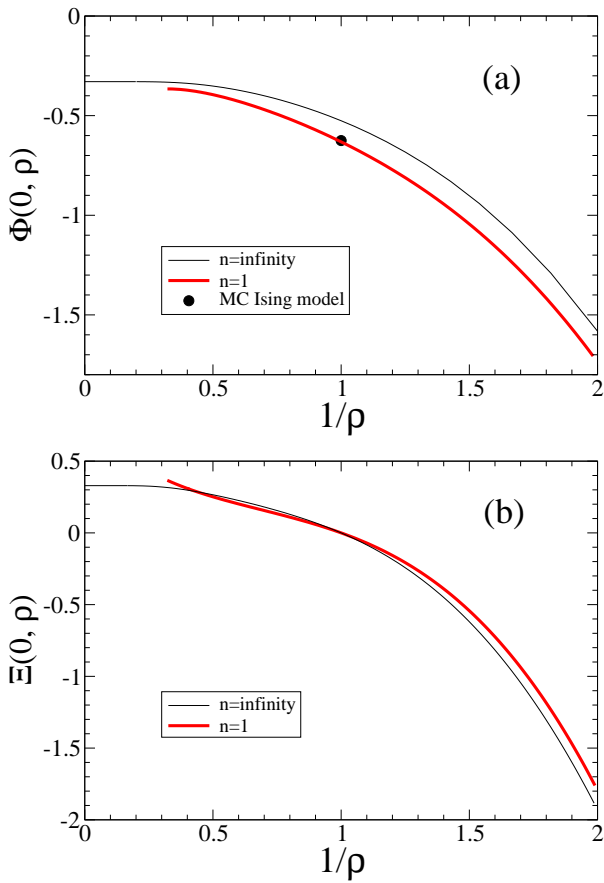

\begin{center}
\subfigure{\includegraphics[clip,width=7.92cm]{FreieEnergieKastenZylinderSphIsingBeiTc.eps}}
\subfigure{\includegraphics[clip,width=7.92cm]{CasimirZYLamplitudeSphIsingBeiTcPhysRevE2010Neu.eps}}
\end{center}
\caption{(Color online)  Critical amplitudes (a) $\Phi(0, \rho)$ (\ref{3r}) and (b) $\Xi(0, \rho)$, (\ref{3rAmpli}), at $T_c$ in three dimensions as a function of the inverse  aspect ratio $1/\rho$ in the
large-$n$ limit [thin lines, from (\ref{3.9x})] and for $n=1$ [thick lines, from (\ref{6-2})]). For $n=\infty$, $\Phi_{cyl}(0)\equiv \Phi(0, \infty) =-0.329= -\Xi_{cyl}(0)\equiv -\Xi(0,\infty)$. At $1/\rho=1$, $\Xi(0, 1)$ vanishes for both $n=\infty$ and $n=1$. Monte Carlo data  of the $d=3$ Ising model for $\rho=1$ by Mon \cite{mon-1} [full circle in (a)]. See also Fig. 8.}
\end{figure}

For film and cylinder geometries in the large - $n$ limit, the low-temperature limits are
\begin{equation}
\label{4ii}
\lim_{\tilde{x} \to -\infty}X_{film}(\tilde{x})= 2 F_{film}^{ex}(-\infty)= {\cal G}_{0,film}(0)= -0.383
\end{equation}
for $d=3$ [see Fig. 3(b)], in  agreement with the earlier result for the spherical model \cite{dan,dan-krech} , and
\begin{equation}
\label{4iiii}
\lim_{\tilde{x_\parallel} \to -\infty}\Xi_{cyl} (\tilde x_\parallel) = -\Phi_{cyl} (-\infty)=-\frac{1}{2}\;{\cal G}_{0,cyl}(0)
\end{equation}
for $2<d<4$, with $-\frac{1}{2}\;{\cal G}_{0,cyl}(0)= 0.719$ for $d=3$ [see Fig. 4(b)].

\subsection*{D. Monotonicity properties at fixed temperature
}
From Figs. 3 and 4 we also infer monotonicity properties at fixed $\tilde x $ and $\tilde x_\parallel $, respectively, i. e., at fixed temperature.
For fixed $\tilde x$,  $F^{ex}(\tilde x, \rho)$ is monotonically decreasing with increasing $\rho$ whereas $X(\tilde x, \rho)$ is monotonically increasing with increasing $\rho$. For fixed $\tilde x_\parallel$,  both $\Phi^{ex} (\tilde x_\parallel, \rho)$ and $\Xi^{ex} (\tilde x_\parallel, \rho)$ are monotonically decreasing with increasing $1/\rho$. This monotonicity is demonstrated by the thin lines in Figs. 5 and 6 at bulk $T_c$ ($\tilde x = 0, \tilde x_\parallel=0$). These lines also exhibit a monotonic change of the curvature towards zero for $\rho \to 0$ and for $1/\rho \to 0$, respectively. For comparison, corresponding curves  are also shown for the $n=1$ universality class (thick lines in Figs. 5 and 6) that will be derived in the subsequent Sections. On the basis of these results we are led to our hypothesis that the monotonicity properties mentioned above are  valid not only for $n=\infty$ but are  general features of the free energy and the Casimir force (for periodic b.c.) that are valid for all $n$ in the whole range $1\leq n \leq \infty$.

\renewcommand{\thesection}{\Roman{section}}
\section*{IV. Perturbation theory for ${\bf n=1}$ in the central finite-size regime }
\renewcommand{\thesection}{\Roman{section}}
\renewcommand{\theequation}{4.\arabic{equation}}
\setcounter{equation}{0}

In this and the subsequent sections we confine ourselves to the case of a one-component order parameter.

\subsection*{A. Perturbation approach for the free energy density}

The basic ingredients of our perturbation approach for $L_{\parallel}^{d-1}\times L $ geometry
are similar to those developed previously for cubic geometry \cite{dohm2008}. The starting point is a decomposition of the variables $\varphi_j = \Phi + \sigma_j$ into the lowest (homogeneous) mode amplitude $\Phi$ and higher-mode contributions $ \sigma_j$,
\be
\label{4.2}
\Phi = V^{-1} \hat \varphi({\bf 0}) = N^{-1} \sum_j \varphi_j,
\ee
\be \sigma_j = \frac{1}{V} {\sum_{\bf k\neq0}} e^{i{\bf k} \cdot
{\bf x}_j} \hat \varphi({\bf k}). \ee
Correspondingly, the Hamiltonian $H$ and the partition function $Z$ are decomposed as
\begin{equation}
\label{4.4}
H = H_0 + \widetilde H (\Phi, \sigma),
\end{equation}
\begin{equation}
H_0 (r_0 \; , u_0 \;, V\; , \Phi^2) = V
\left(\frac{1}{2} r_0 \Phi^2 + u_0 \Phi^4 \right),
\end{equation}
\begin{eqnarray}
\label{4f} \widetilde H(\Phi,\sigma) &=& {\tilde a}^d\Bigg\{\sum_{j=1}^N
\left[( \frac{r_0}{2}+6u_0\Phi'^2) \sigma^2_j +
4u_0\Phi\sigma_j^3 + u_0 \sigma^4_j \right] \nonumber\\ &+&
\sum_{i,j=1}^N \frac{K_{i,j}}{2} (\sigma_i - \sigma_j)^2
\Bigg\},\;
\end{eqnarray}
\be Z = \frac{V^{1/2}}{\tilde a} \int\limits^\infty_{- \infty}  d
\Phi \exp \left\{- \left[H_0 +
{\bare{\Gamma}}(\Phi^2)\right]\right\}, \ee
\begin{equation}
\label{4.8}
{\bare{\Gamma}} (\Phi^2) = -\; \ln \left[\prod_{{\bf
k'\neq0}}\frac{1}{{\tilde a} V^{1/2}} \int \; d \hat \sigma({\bf k})
\right] \exp \left[ - \widetilde H (\Phi, \sigma) \right],
\end{equation}
where $\hat \sigma({\bf k}) \equiv \hat \varphi({\bf k})$ for
${\bf k} \neq{\bf 0}$.
We shall calculate the partition function and the free energy by first determining $ {\bare{\Gamma}} (\Phi^2)$  by means of perturbation theory at given $\Phi$ and subsequently performing the integration over $\Phi$. Since  $\exp [-  {\bare{\Gamma}} (\Phi^2)]$ is  proportional to the
order-parameter distribution function, which is a physical
quantity in its own right, we shall maintain the
exponential form of $\exp [-  {\bare{\Gamma}} (\Phi^2)]$ without further expansion.

The decompositions (\ref{4.4}) - (\ref{4.8}) and the perturbative treatment of the higher modes are reasonable as long as there exists a single lowest mode that is well separated from the higher modes. This is, of course, not the case in the film limit $\rho \to 0$ and in the cylinder limit $\rho \to \infty$ where the system has a lowest mode {\it continuum} and where a revised perturbation approach would be necessary. In Sec. V below, a quantitative estimate will be given to what range of $0 < \rho < \infty$ our perturbation approach is expected to be applicable.

Since the details of the perturbation approach for $f(t, L, L_\parallel )$ are  parallel to those presented in \cite{dohm2008} for cubic geometry  we
directly turn to the result. Our perturbation expression for the bare free energy density reads
\begin{eqnarray}
\label{4z} f = &-& \frac{N-1}{2 V} \ln(2 \pi) + \frac{1}{2}S_0 (r_{0{\rm L}}, L, \rho)\nonumber\\
&-& \frac{1}{V} \ln \int\limits_{-\infty}^\infty d s \;\exp (-
\frac{1}{2} y_0^{eff} s^2 - s^4)   \nonumber\\
&-& \frac{1}{2 V} \ln\; \left[\frac{V^{1/2} w_0^{eff}}
{{\tilde a}^2}\right] - 6 u_0 M_0^2 S_1 (r_{0{\rm L}}, L, \rho) \nonumber\\
&-& 36 u_0^2 M_0^4 S_2 (r_{0{\rm L}}, L, \rho)
\end{eqnarray}
with
\begin{eqnarray}
\label{4z1} y_0^{eff} & = & V^{1/2} u_0^{- 1/2}
\Big\{r_0[1+18u_0S_2(r_{0{\rm L}}, L, \rho)] \nonumber\\&+& 12
u_0S_1(r_{0{\rm L}}, L, \rho)
 +  144 u_0^2M_0^2S_2(r_{0{\rm L}}, L, \rho) \Big\}\;\;\;,
\end{eqnarray}
\begin{equation}
\label{weff} w_0^{eff} \; = \; u_0^{- 1/2}
\left[1+18u_0S_2(r_{0{\rm L}}, L, \rho) \right] \; .
\end{equation}
Here $S_i(r_{0{\rm L}}, L, \rho)$ denote the sums over the higher modes,
\be
\label{S0}
S_0 (r_{0{\rm L}}, L, \rho)= \frac{1}{V} {\sum_{\bf k\neq0}} \ln
\left\{\left[r_{0{\rm
L}} + \delta \widehat K (\mathbf k)\right] {\tilde a}^2\right\},
\ee
\be
\label{4rr} S_m (r_{0{\rm L}}, L, \rho) =  \frac{1}{V} {\sum_{\bf k\neq0}}
\left\{\left[r_{0{\rm L}} + \delta \widehat K({\bf k})\right]
\right\}^{-m}
\ee
for $m=1,2$.
The temperature dependence enters through the parameter
\be
\label{4m} r_{0{\rm L}} (r_0, u_0, V) = r_0 + 12 u_0 M_0^2(r_0, u_0, V)
\ee
as well as through the lowest-mode average
\be
\label{4l} M^2_0 (r_0, u_0, V) = \frac{
\int\limits_{\infty}^\infty d \Phi \;\Phi^2 \exp [-
H_0(r_0, u_0, V)]}{\int\limits_{\infty}^\infty d \Phi \;\exp [- H_0(r_0, u_0, V)]}\;.
\ee
The
positivity of $r_{0{\rm L}} > 0$ for all $r_0$ permits us to
apply the theory to the region below $T_c$. For finite $V$, $M^2_0$ and $r_{0{\rm L}}$ interpolate smoothly
between the mean-field bulk limits above and below $T_c$

\be \lim_{V \rightarrow \infty}   M^2_0 \equiv
 M_{mf}^2 = \left\{ \begin{array}{r@{\quad\quad}l}
                 0 \hspace{1.0cm} & \mbox{for} \; r_0\geq0\;, \\ -r_0/(4u_0) & \mbox{for} \;
                 r_0\leq0\;,
                \end{array} \right.
\ee
and
\be \lim_{V \rightarrow \infty} r_{0{\rm L}} \equiv  r_{mf} = \left\{
\begin{array}{r@{\quad\quad}l}
                 r_0 & \mbox{for} \;\;\; r_0\geq0\;, \\  -2r_0 & \mbox{for} \;\;\;
                 r_0\leq0\;,
                \end{array} \right.
\ee
respectively. In the bulk limit, Eq. (\ref{4z}) correctly contains the bare bulk free energy density in
one-loop order [i.e., up to $O(1)$],
\begin{eqnarray}
\label{4z3} f_b^+  &=&  - \; \frac{\ln (2 \pi)} {2{\tilde a}^d} \; + \;
\frac{1}{2} \int\limits_{\bf k} \ln \{[r_0 + \delta \widehat K
(\mathbf k)] {\tilde a}^2 \} ,\\
\label{4z4} f_b^- &=& \frac{1}{2} r_0 M_{mf}^2 + u_0 M_{mf}^4 -
\; \frac{\ln (2 \pi)} {2{\tilde a}^d} \nonumber\\  &+& \; \frac{1}{2}
\int\limits_{\bf k} \ln \{[- 2 r_0 + \delta \widehat K (\mathbf
k)] {\tilde a}^2\} \;\;\;\;
\end{eqnarray}
above and below $T_c$, respectively. [For the symbol $\int\limits_{\bf k}$ see (\ref{3.3}).]
In the derivation of (\ref{4z}), $\bare{\Gamma} (\Phi^2)$ has been expanded around $ M^2_0$ in powers of $\Phi^2 - M^2_0$ up to $O((\Phi^2 - M^2_0)^2)$. Furthermore, an expansion with respect to $u_0$ at fixed $r_{0{\rm L}}$ has been made and has been truncated such that terms of $O(u^{3/2}_0)$ are neglected.  For a discussion of the order of the
neglected terms see  \cite{dohm2008,dohm1989,EDC}.

\subsection*{B. Dependence on  the aspect ratio ${\bf \rho}$}

Eqs. (\ref{4z}) - (\ref{weff}) are identical in structure with Eqs. (4.26) -(4.28) of \cite{dohm2008} where a cubic geometry was considered. The new point of interest here is the dependence of the bare free energy density $f(t,L_{\parallel}, L)$ on the aspect
ratio  $ \rho = L/L_\parallel$. The $\rho$ dependence enters (i) through the volume
\be
\label{volume1}
V=L^{d-1}_\parallel L = L^d \rho^{1-d},
\ee
(ii) through the lowest-mode average
\be M^2_0 (r_0, u_0, V) = (L^d \rho^{1-d}u_0)^{- 1/2} \; \vartheta_2 (y_0),\ee
\be
y_0  = r_0 (L^d \rho^{1-d}/u_0)^{1/2} \; ,
\ee
\be
\label{theta}
\vartheta_2 (y_0) = \frac{\int\limits_0^\infty d s \;s^2 \exp (-
\frac{1}{2} y_0 s^2 - s^4)} {\int\limits_0^\infty d s \; \exp (-
\frac{1}{2} y_0 s^2 - s^4)} \;,
\ee
and (iii) through the higher-mode sums $S_i(r_{0{\rm L}}, L, \rho)$, (\ref{S0}) and (\ref{4rr}). In the regime of  large $L/{\tilde a}$, large $ L_\parallel/{\tilde a}$, small $0 < (r_{0{\rm L}})^{1/2} {\tilde a} \ll 1$ and fixed $ 0 < L(r_{0{\rm L}})^{1/2}\lesssim O(1)$, $ 0 < L_\parallel (r_{0{\rm L}})^{1/2}\lesssim O(1)$,  these sums are evaluated for finite $0 < \rho < \infty$ in $2<d<4$ dimensions as (see App. B of \cite{dohm2008} and our App. A)
\begin{eqnarray}
\label{4rrp0}&& S_0 (r_{0{\rm L}}, L, \rho)  = \int\limits_{\bf k} \ln \{[ \delta
\widehat K (\mathbf k)] {\tilde a}^2\} +  r_{0{\rm L}} \;
\int\limits_{\bf k} [\delta \widehat K (\mathbf k)]^{-1} \nonumber\\&-&
\frac{2A_d\; (r_{0{\rm L}})^{d / 2}}{d \varepsilon}+ \frac{1}{L^d}\ln
\left(\frac{L^2}{ {\tilde a}^2 4\pi^2}\right) + \frac{1-\rho^{d-1}}{L^d} \ln(r_{0{\rm L}}  {\tilde a}^2) \nonumber\\&+& \;\frac{1}{L^d}
J_0(r_{0{\rm L}} L^2, \rho),\;\;\;\;
\end{eqnarray}
\begin{eqnarray}
\label{4rrp}&& S_1(r_{0{\rm L}}, L, \rho)  =
\; \int\limits_{\bf k}  [ \delta \widehat K (\mathbf k)]^{-1}  \; -
\; \frac{ A_d}{ \varepsilon} \; (r_{0{\rm L}})^{(d-2) /
2}\nonumber\\&+&
\frac{1-\rho^{d-1}}{L^d} (r_{0{\rm L}})^{-1}  +\;\;\frac{(L)^{2-d}}{(4 \pi^2)}
I_1 (r_{0{\rm L}}L^2, \rho),
\end{eqnarray}
\begin{eqnarray}
\label{4rrp} S_2(r_{0{\rm L}}, L, \rho)  &=&
\; \frac{A_d}{2 \varepsilon} \; (d-2)
(r_{0{\rm L}})^{- \varepsilon/2}+
\frac{1-\rho^{d-1}}{L^d} (r_{0{\rm L}})^{-2} \nonumber\\ &+&\;\;\frac{(L)^{4-d}}{(4 \pi^2)^2}
I_2 (r_{0{\rm L}}L^2, \rho)
\end{eqnarray}
with
\begin{eqnarray}
\label
{4.27} J_0(x^2, \rho)= \int\limits_0^\infty
dy y^{-1}
  \Big[\exp {\left[-x^2y/(4\pi^2)\right]}
  \nonumber\\ \times \left\{
  (\pi / y)^{d/2}
  - \;[\rho\; K(\rho^2 y)]^{d-1} \;K(y) + 1 \right\}  -  e^{-y}\Big],\;\;
\end{eqnarray}
\begin{eqnarray}
\label
{4.28} I_m (x^2,\rho) = \int\limits_0^\infty
{\rm{d}}y \;y^{m-1} \exp[- x^2 y / (4 \pi^2)] \nonumber\\
\times \{\;[\rho\; K(\rho^2 y)]^{d-1} \;K(y) - (\pi / y)^{d/2} - 1\}
\qquad
\end{eqnarray}
for $m=1,2$. [For $K(y)$ see (\ref{funktionK}).]

\subsection*{C. Bare perturbation result}

It is appropriate to rewrite the free energy density $f$, (\ref{4z}), in terms of $r_0 - r_{0c}$ where
\begin{equation}
\label{4zzz} r_{0c}   = - \; 12 u_0
\int\limits_{\bf k} [\delta \widehat K ({\bf k})]^{-1}
\end{equation}
is the critical value of $r_0$ up to $O(u_0)$. The resulting function is denoted as $\hat f (r_0 - r_{0c}, u_0, L, \rho, K_{i,j}, \tilde a)$. As we are interested only in the singular part we subtract the non-singular bulk part up to linear
order in $r_0 - r_{0c}$,
\begin{eqnarray}
\label{linear}
f^{(1)}_{ns} (r_0 - r_{0c},  K_{i,j}, \tilde a) &=& -
\frac{\ln (2 \pi)} {2 {\tilde a}^d} \;+ \; \frac{1}{2}
\int\limits_{\bf k} \ln \{[ \delta \hat K ({\bf k})] {\tilde a}^2 \}
\nonumber\\ &+& \frac{ r_0 - r_{0c}}{2} \int\limits_{\bf k}[\delta
\hat K ({\bf k})]^{-1} \;.
\end{eqnarray}
The remaining function
\begin{eqnarray}
\label{deltaf}
 \delta f
(r_0 - r_{0c}, u_0, L, \rho, K_{i,j}, \tilde a) &=& \hat f(r_0 - r_{0c}, u_0, L, \rho,  K_{i,j}, \tilde a) \nonumber\\ &-&
f^{(1)}_{ns} (r_0 - r_{0c}, K_{i,j}, \tilde a)
\end{eqnarray}
has a finite limit for $\tilde a \to
0$ at fixed $r_0 - r_{0c}$ in $2 < d < 4$ dimensions,
\be
\label{delta}\lim_{\tilde a \to 0} \; \delta f (r_0 - r_{0c}, u_0,  L, \rho, K_{i,j}, \tilde a) = \delta f (r_0 - r_{0c}, u_0,
L,  \rho)\;,  \ee
where we have assumed the interaction (\ref{2g}).
(For the justification of taking the  limit $\tilde a \to 0$ see the remarks after Eq. (4.36) of \cite{dohm2008}.) The function (\ref{delta})  still contains a non-singular bulk part  $f^{(2)}_{ns} (r_0 - r_{0c}, u_0)$ proportional to
$(r_0 - r_{0c})^2$. It is convenient to subtract this non-singular bulk part later within the renormalized theory in the asymptotic critical region as described in Subsect. E.
Our perturbation result for the function $\delta f(r_0 - r_{0c}, u_0,
L,  \rho)$  as derived from (\ref{4z}) - (\ref{4l}) and (\ref{volume1}) - (\ref{delta}) reads
\begin{eqnarray}
\label{deltafprime}
&&\delta f(r_0 - r_{0c}, u_0,
L, \rho)=\nonumber\\&&  - \frac{A_d}{ r_{0{\rm L}}^{ \varepsilon/2}} \left[\frac
{r_{0{\rm L}}^2}{4d}  \; + \; \frac{(r_0 - r_{0c})^2} {4
\varepsilon} \; - \; 18 u_0^2 M_0^4\right] \nonumber\\&&+\frac{1}{L^d}\Bigg\{-\rho^{d-1}\ln \int\limits_{-\infty}^\infty d z
\;\exp (- \frac{1}{2} y_0^{eff}(\rho) z^2 - z^4) \nonumber\\&& - \;\frac{1}{2} \ln
\left[\frac{2\pi
w_0^{eff}(\rho)}{L^{\varepsilon/2} \rho^{(d-1)/2}}\right]
+\frac{1}{2} J_0(r_{0{\rm L}} L^2, \rho)\nonumber\\&&  - \frac{3
u_{0} M_0^2 L^2}{2 \pi^2} I_1 (r_{0{\rm L}}L^2, \rho)
- \frac{9 u_0^2 M_0^4
L^4}{4 \pi^4} I_2 (r_{0{\rm L}}L^2, \rho)\Bigg\}\nonumber\\&&
+ \frac{1-\rho^{d-1}}{L^d}\Bigg\{\frac{1}{2} \ln
\left[\frac { w_0^{eff}( \rho)r_{0{\rm L}}
L^2}{L^{\varepsilon/2}\rho^{(d-1)/2}2\pi}\right] - \frac{6
u_{0} M_0^2}{r_{0{\rm L}}}\nonumber\\&&- \frac{36 u_0^2 M_0^4}{ r_{0{\rm L}}^2}\Bigg\},
\end{eqnarray}
\begin{eqnarray}
\label{4bb}&&y_0^{eff}(\rho) = \frac{L^{d/2}\rho^{(1-d)/2}}{u_0^{1/2}}
\Bigg\{ (r_0 - r_{0c})  \Bigg[1 + 18 u_{0}
\Bigg(\frac{A_d(d-2)}{2 \varepsilon \; r_{0{\rm L}}^{
\varepsilon/2}}  \nonumber\\&&+\frac{1-\rho^{d-1}}{L^d\;r_{0{\rm L}}^2}+
\frac{L^\varepsilon}{16\pi^4}I_2(r_{0{\rm L}} L^2, \rho)\Bigg)\Bigg]
\nonumber\\ &&+ \;12u_0\Bigg[- \frac{A_d}{
\varepsilon\;r_{0{\rm L}}^{(2-d)/2}} +
\frac{1-\rho^{d-1}}{L^d\;r_{0{\rm L}}}+
\frac{L^{2-d}}{4\pi^2}I_1 (r_{0{\rm L}} L^2, \rho)\Bigg]\nonumber\\
&&+ \; 144 u_0^2 M_0^2\Bigg[\; \frac{A_d(d-2)}{2
\varepsilon\;r_{0{\rm L}}^{ \varepsilon/2}}
+\frac{1-\rho^{d-1}}{L^d\;r_{0{\rm L}}^2}\nonumber\\&&+
\frac{L^\varepsilon}{16\pi^4}I_2(r_{0{\rm L}} L^2, \rho
)\Bigg]\Bigg\} \;,
\end{eqnarray}
\begin{eqnarray}
\label{4cc}&&w_0^{eff}(\rho)   =  u_0^{- 1/2} \Bigg\{ 1 + 18 u_0
\Bigg[\; \frac{A_d(d-2)}{2 \varepsilon} r_{0{\rm L}}^{-
\varepsilon/2} \nonumber\\ &&+\frac{1-\rho^{d-1}}{L^d}r_{0{\rm
L}}^{-2} +
\frac{L^\varepsilon}{16\pi^4}I_2(r_{0{\rm L}} L^2, \rho
)\Bigg]\Bigg\} \; ,\;\;\;
\end{eqnarray}
where now $r_{0{\rm L}}$ and $M_0^2$ are abbreviations for
\be
r_{0{\rm L}}(r_0 - r_{0c}, u_0, L,\rho)=r_0 - r_{0c} + 12 u_0
M_0^2
\ee
and
\be
M_0^2(r_0 - r_{0c}, u_0, L,\rho)  =  (L^d \rho^{1-d}u_0)^{- 1/2} \; \vartheta_2 (\hat y_0),
\ee
with
\be
\label{haty}
\hat y_0  = (r_0 - r_{0c}) (L^d \rho^{1-d}/u_0)^{1/2} \; .
\ee
Our Eqs. (\ref{deltafprime}) - (\ref{haty}) are applicable to some finite range of $0<\rho<\infty$ and contain Eqs. (4.37) - (4.42) of \cite{dohm2008} as a special case for $\rho = 1$. They are not applicable to the film ($\rho \to 0$) and cylinder ($\rho \to \infty$) limits
below bulk $T_c$.

\subsection*{D. Minimal renormalization at fixed dimension}

As is well known, the bare perturbation form of $\delta f$ requires additive and
multiplicative renormalizations. As the ultraviolet behavior of $\delta f$ does not depend on the aspect ratio $\rho$, the renormalizations are the same as those described in \cite{dohm2008} in terms of the minimal renormalization at fixed dimension $2<d<4$ \cite{dohm1985}. The adequacy of this method in combination with the geometric factor $A_d$, (\ref{4e}), has been demonstrated in \cite{dohm2008} for the case of cubic geometry. Since the aspect ratio $\rho$ is not renormalized we apply the same renormalizations to the present bare expression for $\delta f$, (\ref{deltafprime}). The details are parallel to those in  \cite{dohm2008} which justifies to turn directly to the  renormalized form of $\delta f$. It is defined as
\begin{eqnarray}
\label{5d}   f_R(r, u, L,\rho, \mu) &=& \delta f(Z_rr,
       \mu^{\varepsilon}Z_{u}Z_{\varphi}^{-2}A_d^{-1}u,L, \rho,)\nonumber\\  &-& \frac{1}{8}\mu^{-\varepsilon} r^2 A_d
       A(u,\varepsilon)\;
\end{eqnarray}
where $r$ and $u$ are the renormalized counterparts of $r_0-r_{0c}$ and $u_0$. For the $Z$-factors $Z_i(u,\varepsilon)$ and the additive renormalization constant $A(u,\varepsilon)$ we refer to \cite{dohm2008}. The inverse reference length
$\mu$ is chosen as $\mu^{-1} = \xi_{0 +}$ where $\xi_{0 +} $
is the asymptotic amplitude of the second-moment bulk correlation
length above $T_c$.

The critical behavior is expressed in terms of  a flow parameter $l(t,L,\rho)$ that is determined implicitly by
\be
\label{5v} r_{{\rm L}}(l) = \mu^2l^2.
\ee
The reason for this choice of the flow parameter is given after (\ref{5.8}) below. The dependence of $l$ on $t,L,$ and $\rho$ enters  through the function $r_{{\rm L}}(l)$ which is the renormalized counterpart of $r_{0{\rm L}}$. It is given by
\begin{eqnarray}
\label{5p} r_{{\rm L}}(l) &\equiv& r_{0{\rm L}}(r(l),
l^\varepsilon \mu^\varepsilon
A_d^{-1} u(l), L, \rho) \nonumber\\
&=& r(l) + 12\Big [(\mu l)^{\varepsilon} A_d^{-1} u(l)
L^{- d} \rho^{(d-1)}\Big]^{1/2}\vartheta_2 (y(l)) \;\nonumber\\
\end{eqnarray}
with
\be
\label{5q} y(l) = r (l) \mu^{-2} l^{-2} (L \mu l)^{d/2} \rho^{(1-d)/2}
A_d^{1/2} u(l)^{- 1/2}
\ee
where $\vartheta_2 (y)$ is defined by (\ref{theta}).
The effective renormalized quantities $r(l)$ and $u(l)$ are defined as usual \cite{dohm1985}. Both $r_{{\rm L}}(l)$ and $y(l)$ depend on $t,L,$ and $\rho$. The $t$ dependence originates from  $r(l)$ which depends on $t$ through its initial value  $r(1)=r=at$ with $a=Z_r(u,\varepsilon)^{-1}a_0$.

The effective renormalized counterparts of $
y_0^{eff}(\rho)$ and of $w_0^{eff}(\rho)$  are given by
\begin{eqnarray}
&&y^{eff}(l,\rho) = (l \mu L)^{ d/2} \rho^{(1-d)/2} A_d^{ 1/2}
u(l)^{-1/2}\nonumber\\&& \times \Bigg\{\frac{r (l)}{\mu^2l^2}\Big[1 + 18 u(l)R_2
\Big(\frac{r_{{\rm L}} (l)}{ \mu^2l^2 }, l \mu L,\rho \Big)\Big]
\nonumber\\&&+ 12 u (l) R_1 \Big(\frac{r_{{\rm L}}(l) }{
\mu^2l^2 }, l \mu L, \rho, \Big)
\nonumber\\&&
+ 144  (l \mu L)^{- d/2} \rho^{(d-1)/2} A_d^{- 1/2}
u(l)^{3/2} \vartheta_2 (y(l)) \nonumber\\&& \times R_2 \Big(\frac{r_{{\rm L}}
(l)}{ \mu^2l^2 }, l \mu L, \rho \Big) \Bigg\} \qquad \qquad
\end{eqnarray}
and
\begin{eqnarray}
w^{eff} (l, \rho) \; = \; u (l)^{- 1/2} \Big[1 + 18 u(l)R_2
\Big(\frac{r_{{\rm L}} (l)}{ \mu^2l^2 }, l \mu L, \rho \Big)\Big]\nonumber\\
\end{eqnarray}
where
\begin{eqnarray}
&&R_1 (q, p, \rho) = \varepsilon^{-1} q [1 - q^{- \varepsilon/2}] +
A^{-1}_d (1 - \rho^{d-1}) q^{-1} p^{-d} \nonumber\\
&&\hspace{2.0cm}+ \; p^{\varepsilon - 2} (4 \pi^2 A_d)^{-1} I_1 (q
\;p^2, \rho),
\end{eqnarray}
\begin{eqnarray}
\label{5.8}
&&R_2 (q, p, \rho) = - \; \varepsilon^{-1} [1 - q^{- \varepsilon/2} ]
- \frac{1}{2} q^{- \varepsilon / 2}\nonumber\\
&& + A_d^{-1} (1 - \rho^{d-1})
q^{-2} p^{-d} + \; p^\varepsilon (16 \pi^4 A_d)^{-1} I_2 (q
\;p^2, \rho) \; ,\nonumber\\
\end{eqnarray}
with $I_m$ defined  by (\ref{4.28}).
The dependence of the functions $R_i$ on the ratio $r_{{\rm L}}(l)/ (\mu^2l^2)$ is the reason for the choice (\ref{5v}) of the flow parameter. It  ensures the standard choice in the bulk limit both above and below $T_c$ \cite{dohm1985}
\be
\lim_{L \rightarrow \infty} \lim_{\tilde L \to \infty} \mu^2l^2 =
\left\{
\begin{array}{r@{\quad \quad}l}
                         \mu^2 l_+^2 = r(l_+)& \mbox{for} \;T > T_c ,\\
                         \mu^2 l_-^2 = -2r(l_-) & \mbox{for} \;T <
                 T_c ,
                \end{array} \right.
\ee
and implies $ \mu l \propto L^{- 1}\rho^{(d-1)/d}$
for large finite $V$ at $T=T_c$.

After integration of the renormalization-group equation (see Eqs. (5.6) and (5.7) of \cite{dohm2008}), the renormalized free energy density attains the structure
\begin{eqnarray}
\label{structure}
&&f_R(r,u,L,\rho, \mu) =
f_R \big(r(l),u(l),l\mu,L, \rho\big) \nonumber\\&&+ \;\frac{A_d
r(l)^2}{2(l\mu)^\varepsilon} \int\limits_1^l
B(u(l'))\Big\{\exp\int\limits_l^{l'}\Big[2\zeta_r(u(l'')) -
\varepsilon\Big]\frac{dl''}{l''}\Big\}\frac{dl'}{l'}\;\;\;\;\nonumber\\
\end{eqnarray}
where $B(u)$ and $\zeta_r(u)$ are well known field-theoretic functions of bulk theory \cite{dohm2008,dohm1985}. From (\ref{deltafprime}) and (\ref{5d})  we derive the first term on the right-hand side of (\ref{structure}) as
\begin{eqnarray}
\label{5aa}
&&f_R \big(r(l),u(l),l\mu,L, \rho\big) \nonumber\\&&= -\; A_d (l\mu)^d / (4d) +
18 u(l) L^{-d}\rho^{d-1} \;[\vartheta_2(y(l))]^2 \nonumber\\
&&+\frac{1}{L^d}\Bigg\{\;- \rho^{d-1}\;  \ln \int\limits_{-\infty}^\infty d
z \;\exp \big[- \frac{1}{2}  y^{eff}(l, \rho) z^2 - z^4\big] \nonumber\\&&
-\frac{1}{2}\ln \left[\frac{2 \pi A_d^{1/2} w^{eff} (l,
\rho)}{(l \mu L)^{\varepsilon/2} \rho^{(d-1)/2}}\right]\;
+ \;\frac{1}{2} J_0(l^2 \mu^2 L^2,\rho) \nonumber\\&&- \frac{3(l\mu
L)^{\varepsilon/2} \rho^{(d-1)/2}u(l)^{1/2}}{2\pi^2 A_d^{1/2}}
\;\vartheta_2(y(l))\; I_1(l^2 \mu^2 L^2,\rho) \nonumber\\
&&- \; \frac{9(l\mu L)^\varepsilon \rho^{d-1}u(l)}{4\pi^4 A_d
}\;[\vartheta_2(y(l))]^2\; I_2(l^2 \mu^2 L^2,\rho)\Bigg\}\nonumber\\
&&+ \; \frac{1 - \rho^{d-1}}{ L^d}  \Bigg\{  \frac{1}{2}\ln
\left[\frac{ A_d^{1/2} w^{eff} (l, \rho)l^2 \mu^2 L^2}{(l \mu
L)^{\varepsilon/2} \rho^{(d-1)/2}2 \pi}\right] \nonumber\\&& -6 u
(l)^{1/2} (l \mu L)^{-d/2}\rho^{(d-1)/2}
A_d^{-1/2}\vartheta_2(y(l))\nonumber\\&&-36 u(l) (l \mu
L)^{-d}\rho^{d-1}A_d^{-1} \;[\vartheta_2(y(l))]^2 \Bigg\}.
\end{eqnarray}

\subsection*{E. Finite-size scaling function of the free energy density}

It is straight forward to show that the asymptotic form (\ref{3h}) of the singular part $f_{s}$
of the free energy density is obtained from $f_R$, (\ref{structure}), in the
limit of small $l \ll 1$ or
$ l\rightarrow 0$ as
\be
\label{6k} f_R  \rightarrow f_{s}(t,L,L_\parallel) = L^{-d}
F(\tilde x,\rho)
\ee
with the scaling variable
$\tilde x $, (\ref{3j}).
 In this limit we have $u(l) \rightarrow
\;u(0)\equiv  u^*$, $ r(l)/(\mu^2l^2)\rightarrow \;Q^*\; t
\;l^{-1/\nu}$,
\be
\label{6c} y(l) \rightarrow \; \tilde y = \tilde x \;Q^*\; (\mu
l L)^{- \alpha / (2 \nu)}\rho^{(1-d)/2} A_d^{1/2} {u^*}^{-1/2},
\ee
and $ \mu l
L \rightarrow \; \tilde l = \tilde l(\tilde x,\rho)$ where the
function $\tilde l (\tilde x,\rho)$ is determined implicitly by
\be
\label{6f} \tilde y + 12 \vartheta_2(\tilde y) = \rho^{(1-d)/2}
\tilde l^{d/2} A_d^{1/2} {u^*}^{-1/2},
\ee
\be
\label{6g} \tilde y =\; \tilde x\;Q^*\;\tilde l^{- \alpha / (2
\nu)}\rho^{(1-d)/2} A_d^{1/2} {u^*}^{-1/2}.
\ee
These two equations also determine $\tilde y = \tilde
y(\tilde x,\rho)$. In (\ref{6c})- (\ref{6g}) we have used the hyperscaling relation
$ 2 - \alpha = d \nu $.
The factor $Q^* =Q (1, u^*, d)$ is the fixed point value of the amplitude function $Q (1, u, d)$
of the bulk correlation length above $T_c$ \cite{dohm2008,dohm1985,krause,EDC}.
Furthermore we have, in the small - $l$ limit,
\be
\label{6h} w^{eff}(l, \rho )\rightarrow \; W (\tilde x, \rho) = {u^*}^{-
1/2} \left[1 + 18 \; u^* R_2(1, \tilde l, \rho) \right] \;,
\ee
\begin{eqnarray}
\label{6i} && y^{eff}(l, \rho)\rightarrow \; Y(\tilde x, \rho
) = {\tilde l}^{ d/2}\rho^{(1-d)/2} A_d^{ 1/2} {u^*}^{-1/2}\nonumber\\
&&\times \Bigg\{Q^* \tilde x \;{\tilde l}^{-1/\nu}\Big[1 + 18
u^*R_2(1,\tilde l,\rho) \Big] + \;12 u^* R_1(1, \tilde l,\rho )
\nonumber\\ && + \; 144 \Big[{u^*}^{3} {\tilde l}^{- d}\rho^{(d-1)}
A_d^{- 1}\Big]^{1/2}
 \vartheta_2 (\tilde y )R_2(1,\tilde l, \rho)  \Bigg\}.
\end{eqnarray}
For $ l \ll 1$, the last integral term in
(\ref{structure}) contains both a contribution $\propto  t^2 l
^{-\alpha/\nu} $ to the singular finite-size part $f_{s}(t,L,L_\parallel)$ and a
contribution $\propto t^2$ to the nonsingular bulk part
$f^{(2)}_{ns,b} $ of $\delta f$  mentioned after  (\ref{delta}) (see also the comment on Eq. (6.8) of \cite{dohm2008}). This nonsingular part will be neglected in the following.

Eqs. (\ref{structure}), (\ref{5aa}), and (\ref{6k})-(\ref{6i}) lead to the finite-size scaling function
\begin{eqnarray}
\label{6l} &&F(\tilde x, \rho)= -\; A_d \; \left[
\frac{\tilde l^d}{4d} \; + \;\frac{\nu\;{Q^*}^2 \tilde x^2 \tilde
l^{- \alpha/\nu}}{2\alpha} \;B(u^*)\right] \nonumber\\ && + \; 18
u^*\rho^{d-1} \left[\vartheta_2 (\tilde y)\right]^2 - \frac{1}{2}\ln
\left(\frac{2 \pi A_d^{1/2} W (\tilde x, \rho)}{
\tilde l^{\varepsilon/2}\rho^{(d-1)/2}}\right) \nonumber\\
&&-\;\rho^{d-1}  \ln \int\limits_{-\infty}^\infty d z \;\exp \big[-
\frac{1}{2}  Y(\tilde x, \rho) z^2 - z^4\big] \nonumber\\
&&+ \frac{1}{2}J_0( {\tilde l}^2, \rho) - \frac{3\;{\tilde
l}^{\varepsilon/2} {u^*}^{1/2}\rho^{(d-1)/2}}{2\pi^2 A_d^{1/2}}
\;\vartheta_2(\tilde y)\; I_1( {\tilde l}^2, \rho) \nonumber\\ &&-
\frac{9\;{\tilde l}^\varepsilon u^*\rho^{d-1}}{4\pi^4 A_d }
\;[\vartheta_2(\tilde y)]^2\; I_2( {\tilde l}^2, \rho) \nonumber\\&&+
\; (1 - \rho^{d-1})  \Bigg\{
\frac{1}{2}\ln \left[\frac{ A_d^{1/2} W (\tilde x,
\rho){\tilde l}^{d/2}}{2 \pi \rho^{(d-1)/2}}\right] \nonumber\\&& -6
{u^*}^{1/2} \tilde l^{-d/2}\rho^{(d-1)/2}
A_d^{-1/2}\vartheta_2(\tilde y)\nonumber\\&&-36 u^* \tilde
l^{-d}\rho^{d-1}A_d^{-1} \;[\vartheta_2(\tilde y)]^2 \Bigg\}
\end{eqnarray}
with $A_d, J_0,I_m$ and $\vartheta_2$ defined in (\ref{4e}), (\ref{4.27}), (\ref{4.28}), and (\ref{theta}), respectively.
Eq. (\ref{6l}) is the central analytic result of the present paper for the case $n=1$. It is valid for $2<d<4$ in the central finite-size regime (between the dahed lines of Fig. 2), i.e., in the range $L\gg \tilde a$,  $L_{\parallel}\gg \tilde a$  and
$0 \leq |\tilde x| \lesssim O (1)$ above, at, and below $T_c$ for finite $\rho$.  For the special case $\rho=1$, Eq. (\ref{6l}) is identical with Eq. (6.10) of  \cite{dohm2008}. For $d=3$, Eq. (\ref{6l})  reduces to Eq. (9) presented in \cite{dohm2009}. It incorporates the correct bulk critical exponents $\alpha$ and $
\nu$ and the complete bulk function $B(u^*)$ (not only in one-loop
order). The Borel resummed values of the fixed point value $u^*$ \cite{larin}, of $B(u^*)$ \cite{larin}, and of $Q^*$ \cite{dohm1985,krause,EDC} in three dimensions are given after Eq. (\ref{6.15}) below. There is  only one adjustable parameter that is contained
in the nonuniversal bulk amplitude $\xi_{0+}$ of the scaling
variable $\tilde x$. For finite $L$ and $L_\parallel$, $f_s (t, L,L_\parallel)$ is
an analytic function of $t$ near $t = 0$, i.e.,  $F(\tilde x, \rho)$  is an analytic function of  $\tilde x$ near $\tilde x=0$ at finite $\rho$, in agreement with
general analyticity requirements.

The bulk part $F_b^\pm(\tilde x)$ of $F(\tilde x, \rho)$ is obtained from (\ref{6l}) in the large - $|\tilde x|$ limit. It  is represented by  (\ref{3jj}), with the universal bulk amplitude ratios
\be
\label{6o}  Q_1 = \; - \; A_d
Q^{* d \nu} \left[\frac{1}{4d} \; + \; \frac{\nu}{2 \alpha} \; B
(u^*) \right] \; ,
\ee
\begin{eqnarray}
\label{6p} &&
\frac{A^-}{A^+} \nonumber\\ && = \; 2^{d \nu} \; \frac {1 / (64
u^*) \; + 1/(4d) \; + 81 u^*/64 \; + \; \nu B (u^*) / (8 \alpha) }
{1/(4d) \; + \; \nu B (u^*) / (2 \alpha)} \; \nonumber\\
\end{eqnarray}
given by Eqs. (6.19) and (6.20) of \cite{dohm2008}. We then obtain from (\ref{6l}) the scaling function $F^{ex} (\tilde x, \rho)$, (\ref{3kk}), of the excess free energy density
which determines the scaling function $X(\tilde x,\rho)$ of the Casimir force according to (\ref{3nn}). By definition, the functions $F^{ex} (\tilde x, \rho)$ and $X(\tilde x,\rho)$ have a weak singularity at $\tilde x=0$ arising from the subtraction of the bulk term $F_b^\pm(\tilde x)$.

\section*{V. Quantitative results in three dimensions in the central finite-size regime}
\renewcommand{\thesection}{\Roman{section}}
\renewcommand{\theequation}{5.\arabic{equation}}
\setcounter{equation}{0}

\subsection*{A. Amplitudes at $T_c$ and monotonicity hypothesis}
Of particular interest is the finite-size amplitude at $T_c \;,$
\begin{eqnarray}
\label{6-2} &&F(0,\rho) = \left(18 - \frac{36}{d}\right)
u^* \rho^{d-1}\left[\vartheta_2 (0)\right]^2 \nonumber\\&-& \frac{1}{2}\ln
\left(\frac{2 \pi A_d^{1/2} W_c
(\rho)}{ \tilde l_c^{\varepsilon/2}\rho^{(d-1)/2}}\right) \nonumber\\
&-& \rho^{d-1} \ln \int\limits_{-\infty}^\infty d z \;\exp
\big[- \frac{1}{2}  Y_c(\rho) z^2 - z^4\big] \nonumber\\
&+& \frac{1}{2} \; J_0( {\tilde l_c}^2,\rho)  - \frac{\tilde
l_c^2}{8 \pi^2} \; I_1( {\tilde l_c}^2, \rho) -
\frac{\tilde l_c^4}{64 \pi^4} \; I_2( {\tilde l_c}^2, \rho)
\nonumber\\&& + \; (1 - \rho^{d-1})  \Bigg\{ \frac{1}{2}\ln \left[\frac{
A_d^{1/2} W_c (\rho){\tilde l_c}^{d/2}}{2 \pi \rho^{(d-1)/2}}\right] -
\frac{3}{4} \Bigg\}
\end{eqnarray}
where $\tilde l_c^{d/2} =12 {u^*}^{1/2} \rho^{(d-1)/2} A_d^{-1/2}
\vartheta_2(0)$ and
\be
\label{6-3}W(0,\rho)\equiv W_c (\rho) = {u^*}^{- 1/2} \left[1 + 18 \;
u^* R_2(1, \tilde l_c, \rho) \right],
\ee
\be
\label{6-4}Y(0,\rho)\equiv Y_c(\rho) = 144 u^* \vartheta_2 (0) \Bigg\{
R_1(1, \tilde l_c, \rho) + \; R_2(1,\tilde l_c,\rho) \Bigg\}
\ee
with $\vartheta_2 (0) \;=\; \Gamma (3/4) / \Gamma (1/4) $ and
\be
\label{6-6}R_1(1, \tilde l_c, \rho )= \frac{\tilde l_c^{2-d}}{4 \pi^2
A_d} I_1( {\tilde l_c}^2,
\rho)+A_d^{-1}(1-\rho^{d-1})\tilde{l}_c^{-d} \; ,
\ee
\be
\label{6.15}R_2(1,\tilde l_c, \rho)= - \frac{1}{2} + \frac{\tilde
l_c^\varepsilon}{16 \pi^4 A_d}  I_2( {\tilde l_c}^2, \rho
)+A_d^{-1}(1-\rho^{d-1})\tilde{l}_c^{-d} \; .
\ee
\begin{figure}[!ht]
\begin{center}
\subfigure{\includegraphics[clip,width=7.92cm]{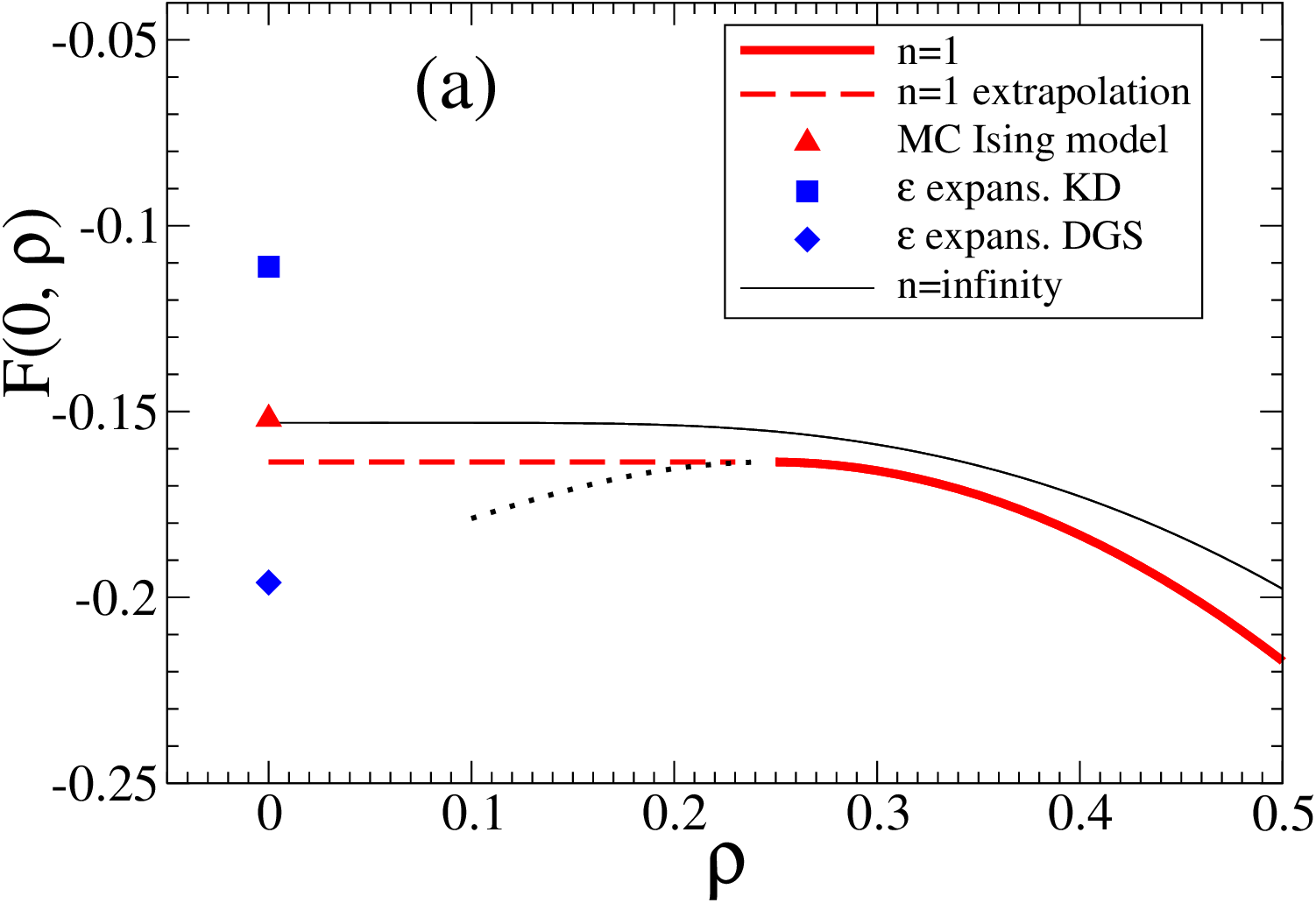}}
\subfigure{\includegraphics[clip,width=7.92cm]{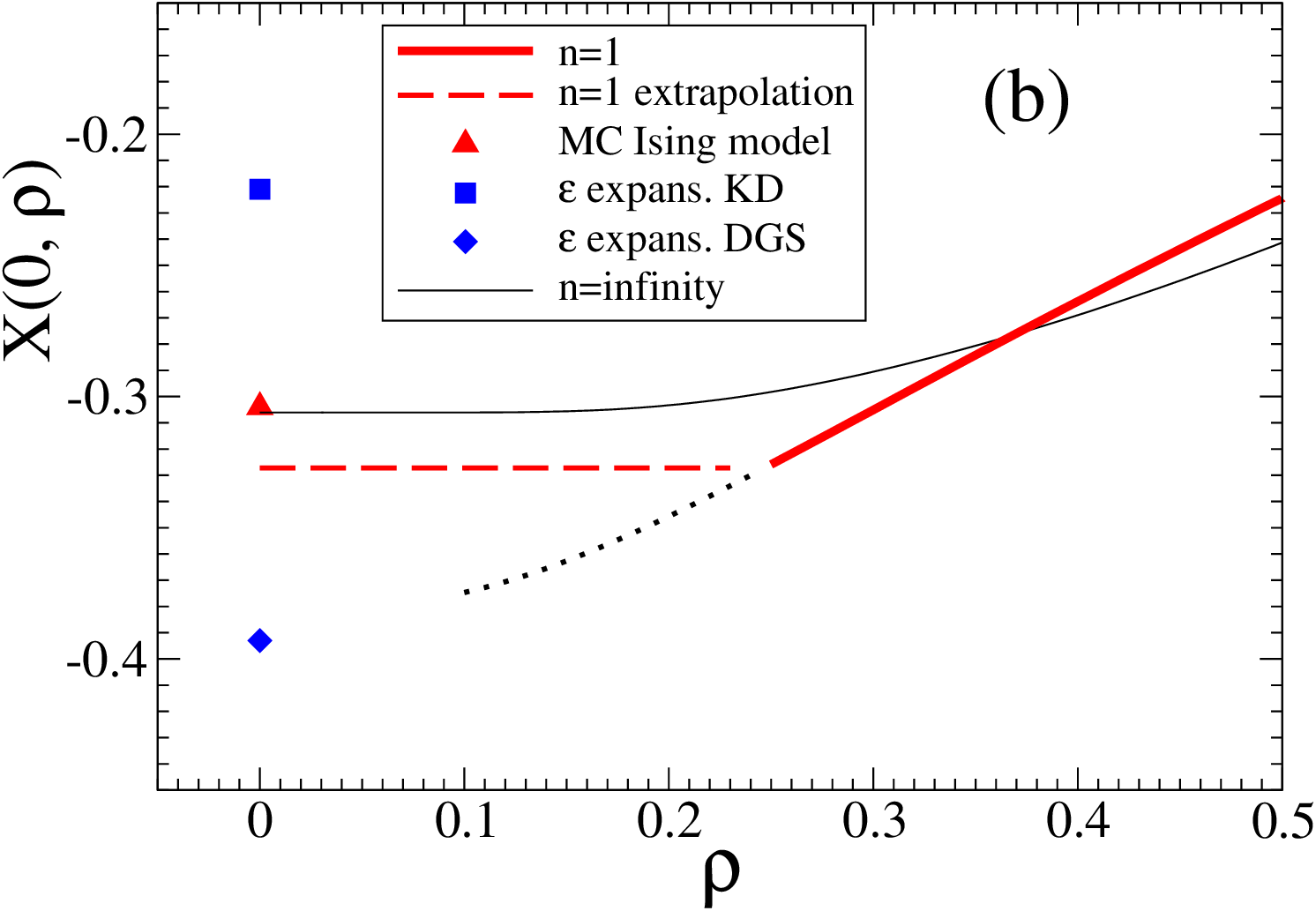}}
\end{center}
\caption{(Color online) Critical amplitudes (a)  $F(0, \rho)$, (\ref{3kk}),  and (b) $X(0, \rho)$, (\ref{3nAmplitude}), at $T_c$ in three dimensions as a function of the aspect ratio $\rho$ for $n=1$ [thick lines, from (\ref{6-2})]) and in the
large-$n$ limit
[thin lines, from (\ref{3.9x})]. The maximum $-0.1636$ of the $n=1$ line in (a)
is at $\rho_{max}=0.2470$. The dashed lines are the extrapolations of the $n=1$ lines from $\rho=\rho_{max}$ to
$\rho=0$ corresponding to film geometry. The dotted lines
represent (\ref{6-2})
in the regime $\rho < \rho_{max}$ where our perturbation theory is not applicable. MC estimate for the d=3 Ising model  from \cite{vasilyev2009} (triangles), $\varepsilon$ expansion results for $n=1$  from \cite{KrDi92a} (squares) and from \cite{DiGrSh06} (diamonds). See also Fig. 5.}
\end{figure}
\begin{figure}[!ht]
\begin{center}
\subfigure{\includegraphics[clip,width=7.92cm]{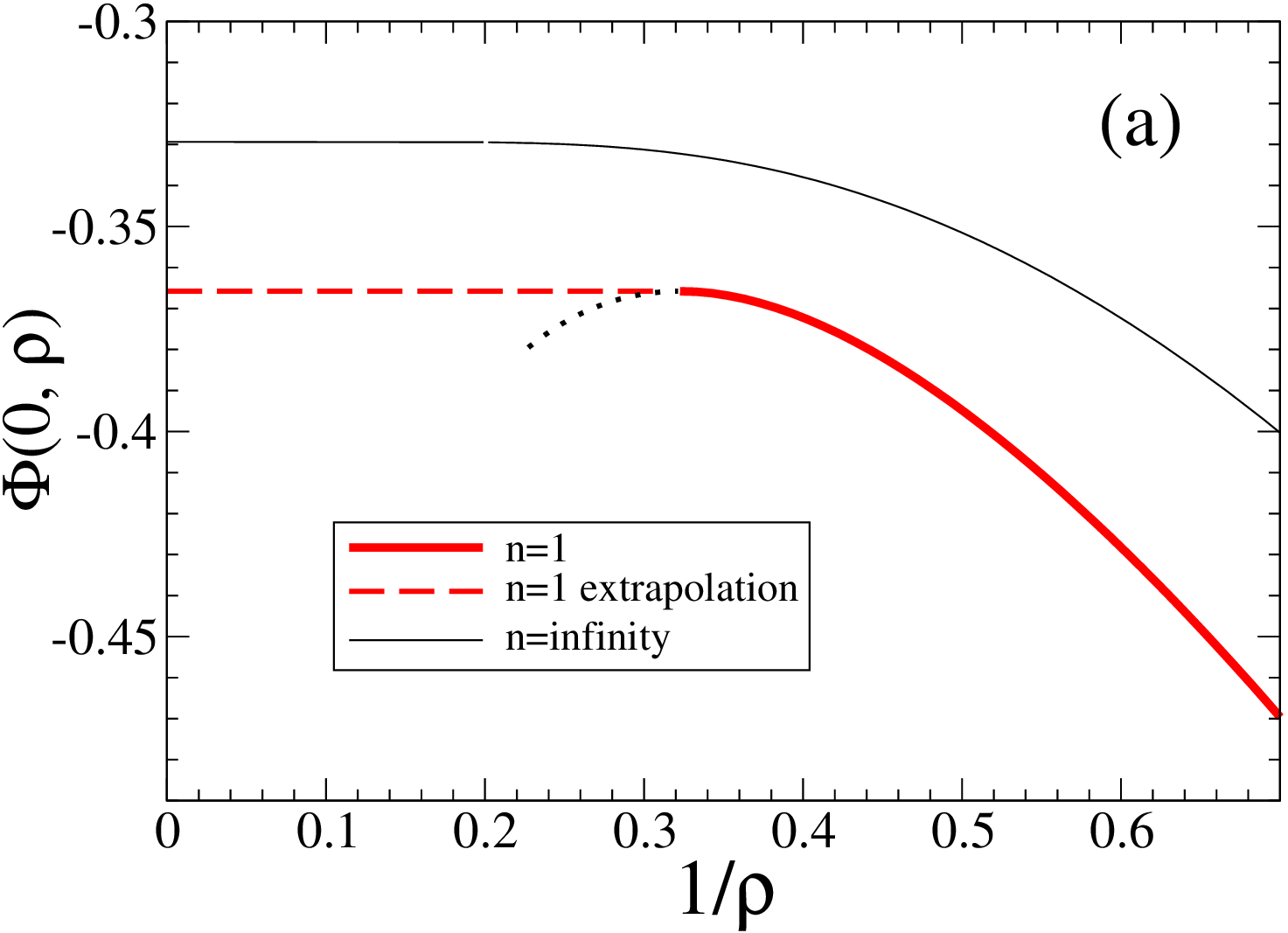}}
\subfigure{\includegraphics[clip,width=7.92cm]{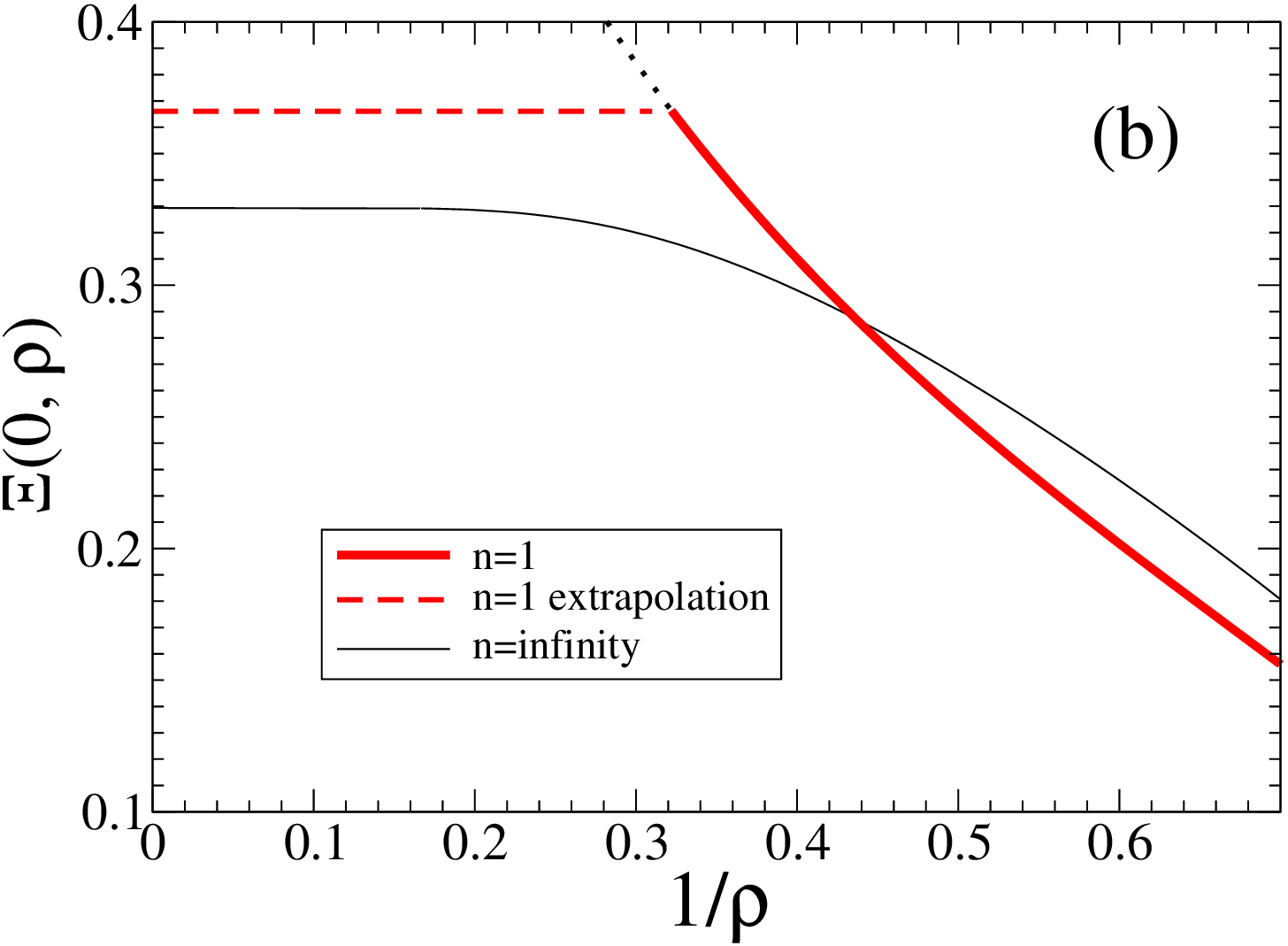}}
\end{center}
\caption{(Color online) Critical amplitudes (a)  $\Phi(0, \rho)$, (\ref{3r}),  and (b) $\Xi(0,\rho)$,  (\ref{3rAmpli}), at $T_c$ in three dimensions as a function of the inverse aspect ratio $1/\rho$ for $n=1$ [thick lines, from (\ref{6-2})] and in the
large-$n$ limit
[thin lines, from (\ref{3.9x})]. The maximum $-0.3658$ of the $n=1$ line in (a)
is at $(1/\rho)_{max}=0.3223$. The dashed lines are the extrapolations of the $n=1$ lines from $(1/\rho)_{max}$ to
$1/\rho=0$ corresponding to cylinder  geometry. The dotted lines
represent (\ref{6-2})
in the regime of small $1/\rho <  (1/\rho)_{max}$ where our perturbation theory is not applicable. See also Fig. 6.}
\end{figure}
For the application to three dimensions we shall employ the following numerical
values \cite{dohm2008,EDC,liu}: $A_3 = (4\pi)^{-1}$, $\nu=
0.6335$,  $u^*= 0.0412$, $Q^*= 0.945$, $B(u^*) = 0.50$, and $\alpha = 2 - 3 \nu = 0.0995$.
At $T_c$ in three dimensions, the $\rho$ dependence of the flow parameter is given by
\be
\tilde l_c =[12(4\pi u^*)^{1/2}
\Gamma (3/4) / \Gamma (1/4)]^{2/3}\rho^{2/3}= 2.042\; \rho^{2/3}.
\ee
The $\rho$ dependence of the integrals $J_0(\tilde {l_c}^2,\rho)$,
(\ref{4.27}), and $I_m(\tilde l_c^2,\rho)$, (\ref{4.28}), needs to
be computed numerically. The resulting amplitudes $F (0,\rho) $,  $X(0,\rho)$, $\Phi(0,\rho)$, and $\Xi(0,\rho)$ as determined by (\ref{6-2}),  (\ref{3nAmplitude}), (\ref{3r}), and (\ref{3rAmpli}) are shown by the thick lines  in Figs. 5 and 6, respectively, in a finite range of $ \rho$ and $1/\rho$. At $\rho=1$, perfect agreement with the MC data by Mon \cite{mon-1} [full circle in Figs. 5(a) and 6(a)] is found.

Figs. 5 and 6 demonstrate the weakness of the $n$ dependence at $T_c$. On the basis of the monotonicity of the curves for the case $n=\infty$ (thin curves in Figs. 5 and 6) we  expect monotonicity also for the $n=1$ curves.  As shown in the magnified plots of Figs. 7 (a) and 8 (a), $F(0,\rho) $ and $\Phi(0,\rho)$ indeed have the expected monotonic behavior, but only in the restricted range $\rho \geq \rho_{max}=0.2470$ and $1/\rho \geq (1/\rho)_{max} = 0.3223$, respectively. As expected on general grounds, the lowest-mode separation approach should fail for sufficiently small $\rho < \rho_{max}$ or $1/\rho < (1/\rho)_{max}$, respectively, near the film and the cylinder limit (dotted lines in Figs. 7 and 8) where the higher modes are no longer well separated from the single lowest mode. Thus our hypothesis of monotonicity provides the following quantitative estimate for the range of the aspect ratio $\rho$ within which our lowest-mode separation approach for the free energy is expected to be reliable:
\begin{equation}
1/4 \lesssim \rho \lesssim 3.
\end{equation}
Furthermore we expect a negligible dependence on $\rho$ and $1/\rho$ in the range $\rho < \rho_{max}$ or $1/\rho < (1/\rho)_{max}$, respectively, corresponding to the extrapolations (dashed lines) in Figs. 7(a) and 8(a). This leads to our prediction of the $n=1$ amplitudes of the scaling functions of the excess free energy density at bulk $T_c $ for the film and for the cylinder in three dimensions:
\begin{eqnarray}
\label{filmampli}
F_{film}(0)\approx F(0,\rho=1/4)=-0.164,\\
\label{cylampli}
\Phi_{cyl}(0)\approx \Phi(0,\rho=3)=-0.366.
\end{eqnarray}
The corresponding results for the Casimir amplitudes $X(0,\rho)$ and $\Xi(0, \rho)$ are shown in  Figs. 7 (b) and 8 (b); they follow from those of $F(0,\rho) $ and $\Phi(0,\rho)$ by means of the exact relations [compare (\ref{3nn}) and (\ref{3rrr})]
\begin{equation}
\label{3nAmplitude}  X(0,\rho) =(d-1)  F(0,\rho) -
\rho
\frac{\partial F(0,\rho)}{\partial \rho},
\end{equation}
\begin{eqnarray}
\label{3rAmpli}\Xi(0, \rho)&=&
-\Phi(0, \rho) \;+ \;(1/\rho)\;
\frac{{\partial \Phi(0,\rho)}}{\partial (1/\rho)}.\;\;
\end{eqnarray}
From (\ref{filmampli}) and (\ref{cylampli}) we obtain our prediction of the $n=1$ amplitudes of the Casimir force scaling functions at bulk $T_c $ for the film and for the cylinder in three dimensions [dashed lines in Figs. 7 (b) and 8 (b)]:
\begin{eqnarray}
\label{Xfilmampli}
X_{film}(0)\equiv X(0,0)= 2F_{film}(0)=-0.328,
\\\label{Xicylampli}\Xi_{cyl}(0)\equiv \Xi(0,\infty)=-\Phi_{cyl}(0)=0.366.
\end{eqnarray}
Our results for $F_{film}(0)$ and $X_{film}(0)$ are in good agreement with the MC estimates \cite{vasilyev2009} $\Delta_P=-0.152$ and $2\Delta_P=-0.304$ [triangles in Fig. 7] for the three-dimensional Ising model in film geometry at bulk $T_c$. The previous $\varepsilon$ expansion results  up to $O(\varepsilon)$ \cite{KrDi92a}
[squares in Fig. 7], and up to $O(\varepsilon^{3/2})$ \cite{DiGrSh06}
[diamonds in Fig. 7], are in less good agreement  with the MC estimates.

It would be interesting to test our predictions for $\Phi_{cyl}(0)$, (\ref{cylampli}), and $\Xi_{cyl}(0)$, (\ref{Xicylampli}), by MC simulations for the three-dimensional Ising model in cylinder geometry.

\subsection*{B. Finite-size scaling functions }
Now we turn to a discussion of the temperature dependence.
In Figs. 9 and 10 we show the scaling functions  $F^{ex}(\tilde x, \rho)$, $X(\tilde x, \rho)$, $\Phi^{ex}(\tilde x_{\parallel}, \rho)$,  and  $\Xi(\tilde x_{\parallel}, \rho)$ for $n=1$ in three dimensions
for slab, cube, and rod geometries, respectively, with finite aspect ratios in the range   $1/4 \leq \rho \leq 5/2$ , as derived from (\ref{6l}), (\ref{3kk}),   (\ref{3nn}), (\ref{3rr}), and (\ref{3rrr}). It is expected that these curves are applicable to the central finite-size regime $|\tilde x| \lesssim O(1)$  and $|\tilde x_{\parallel}| \lesssim O(1)$ but not to $|\tilde x| \gg 1$ and $|\tilde x_{\parallel}| \gg 1$. (For a more precise estimate see below.) Figs. 9 and 10 should be compared with the corresponding Figs. 3 and 4 for the case $n=\infty$.

%
\begin{figure}[!ht]
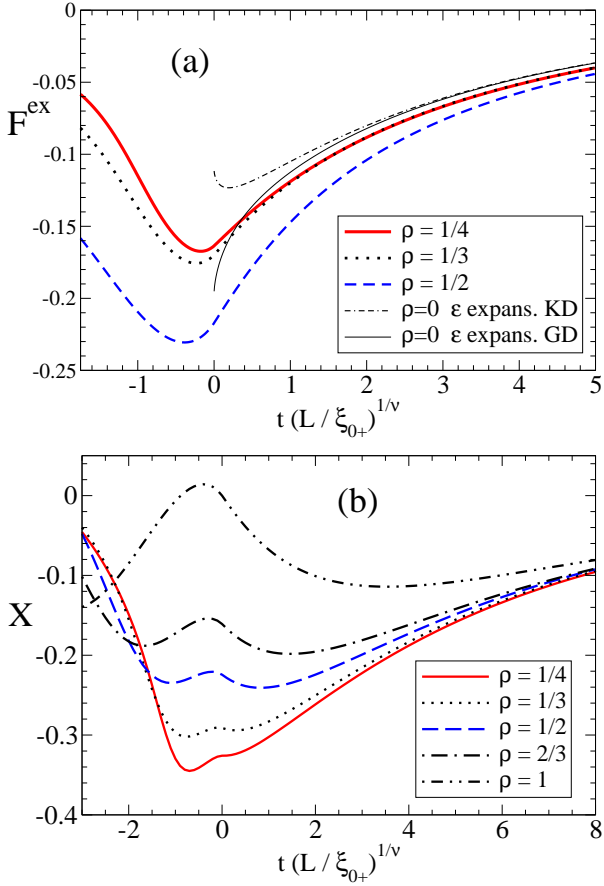

\begin{center}
\subfigure{\includegraphics[clip,width=7.92cm]{ExcessFreieEnergieIsingAlleTShapeEinhalbBisEinviertelundMCundEpsilonPhysRevE2010.eps}}
\subfigure{\includegraphics[clip,width=7.92cm]{CasimirSkalenfktngleich1ShapeEinViertelBisEinsDick.eps}}
\end{center}
\caption{(Color online) Scaling function (a) $F^{ex}(\tilde x, \rho)$, (\ref{3kk}), (\ref{6l}),   and (b) $X(\tilde x, \rho)$, (\ref{3nn}), (\ref{3kk}), (\ref{6l}), as a function of  $\tilde x = t (L/\xi_{0+})^{1/\nu}$ for $n=1$ in three dimensions
for slab geometry with finite aspect ratio  $\rho= 1/4$ (solid lines),  $\rho= 1/3$ (dotted lines),  $\rho= 1/2$ (dashed lines), $\rho= 2/3$ (dot-dashed line), $\rho=1$ (double-dot-dashed line).  Thin lines in (a): $\varepsilon$ expansion results for $\rho=0$ from \cite{KrDi92a,GrDi07}. }
\end{figure}
%

We see that there are significant differences between the cases $n=1$ and $n=\infty$. Figs. 9 (a) and 10(a) exhibit a nonmonotonicity of  $F^{ex}(\tilde x, \rho)$ and $\Phi^{ex}(\tilde x_{\parallel}, \rho)$ for $n=1$ with minima slightly below $T_c$ for all $\rho$. Such minima should also persist in the $n=1$ film ($\rho=0$) system and in the $n=1$ cylinder ($1/\rho=0$) system whose scaling functions  should be close to our curves for $\rho = 1/4$ and $1/\rho =2/5$, respectively. There is no good agreement at $T_c$ between our $\rho = 1/4$ curve in Fig. 9 (a) and the $\varepsilon$ expansion results (thin lines ) of \cite{KrDi92a,GrDi07} for $\rho=0$. The latter exhibit an unphysical singularity at $\tilde x = 0$ (i.e., at bulk $T_c$) that arises from the $\varepsilon$ expansion results \cite{KrDi92a,GrDi07} for the term $F(\tilde x, \rho)$ in (\ref{3kk}) which should be an analytic function of $\tilde x $ near $\tilde x = 0$ since the film transition occurs at a distinct temperature $T_{c,film}$ below bulk $T_c$. Our curves contain a different type of singularity at $\tilde x = 0$ that arises from subtracting the singular bulk part $F_b^\pm(\tilde x)$ in (\ref{3kk}); this singularity is very weak and not visible in Figs. 9 and 10.

In Fig. 9 (b) our results show an unexpected structure of the Casimir force scaling function $X$ near bulk $T_c$ where local {\it maxima} occur with increasing $\rho > 1/4 $. The small shoulder for $\rho = 1/4 $ was already noticed previously \cite{dohm2009}. This structure with local maxima does not exist for $n=\infty$.  Such maxima also persist in the regime of $\rho > 1$ as shown in Fig. 10 (b). As a special feature of the case $\rho=1$, $X$ and $\Xi$ vanish at bulk $T_c$ in three dimensions, as shown by the double-dot-dashed curves in Figs. 9 (b) and 10 (b) [see also Figs. 5 and 6]. In addition, the Casimir force for $n=1$ in a cube changes  sign  at $\tilde x = - 0.884$ and is negative for $\tilde x < -0.884$, contrary to the case $n=\infty$ below $T_c$ [Figs. 3 (b) and 4 (b) for $\rho=1$]. Thus our theory predicts that, in a cube, there is only a small positive region between $\tilde x = - 0.884$ and $\tilde x = 0$.

\begin{figure}[!ht]
\begin{center}
\subfigure{\includegraphics[clip,width=7.92cm]{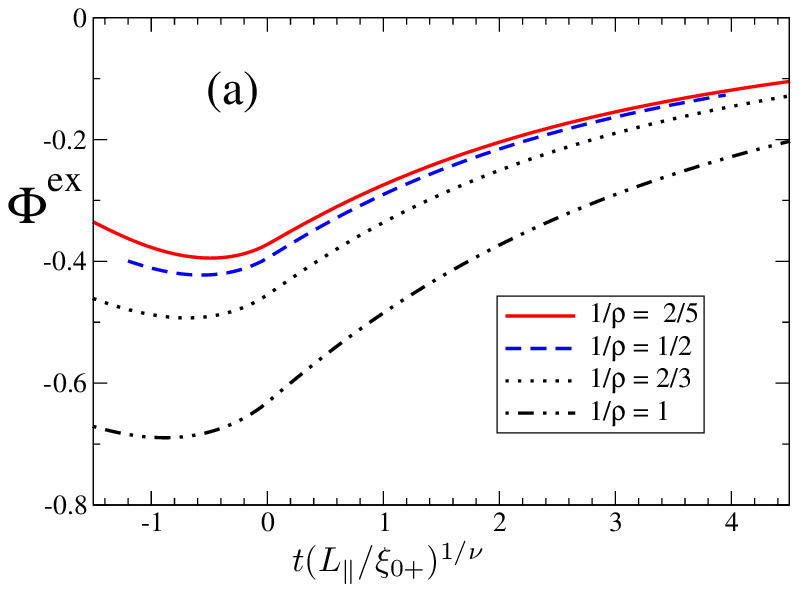}}
\subfigure{\includegraphics[clip,width=7.92cm]{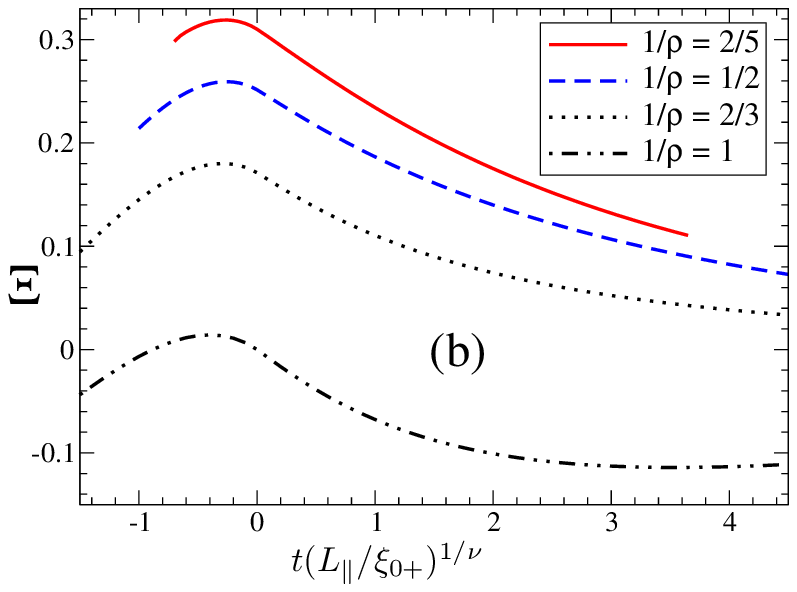}}
\end{center}
\caption{(Color online) (a) Scaling function (a) $\Phi^{ex}(\tilde x_{\parallel}, \rho)$, (\ref{3rr}), (\ref{3kk}), (\ref{6l}),  and (b) $\Xi(\tilde x_{\parallel}, \rho)$, (\ref{3rrr}), as a function of  $\tilde x_{\parallel} = t (L_{\parallel}/\xi_{0+})^{1/\nu}$ for $n=1$ in three dimensions
for rod geometry with finite aspect ratio  $1/\rho= 1$ (double-dot-dashed lines),  $1/\rho= 2/3$ (dotted lines),  $1/\rho= 1/2$ (dashed lines), $1/\rho= 2/5$ (solid lines).  }
\end{figure}

On purely theoretical grounds, it is difficult to provide a precise quantitative estimate for the range of validity of our perturbation approach with regard to the dependence on the scaling variable $\tilde x$. Valuable information, however, has been made available to us by Hasenbusch \cite{hasenbusch} who performed MC simulations for the free energy density of the three-dimensional Ising model in a cubic geometry. These data are shown in Fig. 11, together with our theoretical curve derived from (\ref{3kk}) and (\ref{6l}).
We see that there is good agreement in the range $-0.05 \lesssim \tilde x \lesssim 3$ but significant deviations exist well below $T_c$; small but systematic deviations exist also well above $T_c$. In particular, our perturbation result for $F^{ex}$ has an algebraic approach to a {\it finite} limit $F^{ex}(\infty, \rho)$ for $\tilde x \to \infty$ whereas there should be an exponential decay towards zero (see Sec. VI). From this comparison it is obvious that the lowest-mode separation approach needs to be complemented by a perturbation approach that is valid outside the central finite-size regime. Such an approach will be presented in the subsequent section.

Additional valuable information comes from a comparison of our Casimir force scaling function   with earlier MC data for periodic b.c. in the small - $\rho$ regime  \cite{vasilyev2009}. We recall that the lower limit of applicability of our calculation is $\rho=1/4$ and that the Casimir forces should depend only weakly on $\rho$ for  $\rho < 1/4$, thus it is reasonable to compare our result for $\rho=1/4$ with MC data for $\rho=1/6$ \cite{vasilyev2009}. This comparison is shown in Fig. 12 . Also shown are the previous $\varepsilon$ expansion results for $\rho=0$ from \cite{KrDi92a,GrDi07} which exhibit the same kind of singularity at $\tilde x = 0$ as in Fig. 9 (a). We see good agreement of the MC data with our fixed $d$ perturbation theory in the whole range $-2 \lesssim \tilde x \lesssim 20$. There are systematic deviations only for $\tilde x < -2$ which are less pronounced than those for $F^{ex}$ in the same region. In the subsequent section we shall explain this different degree of agreement between our  theory and the MC data shown in Figs. 11 and 12.

\begin{figure}[!h]
\includegraphics[clip,width=80mm]{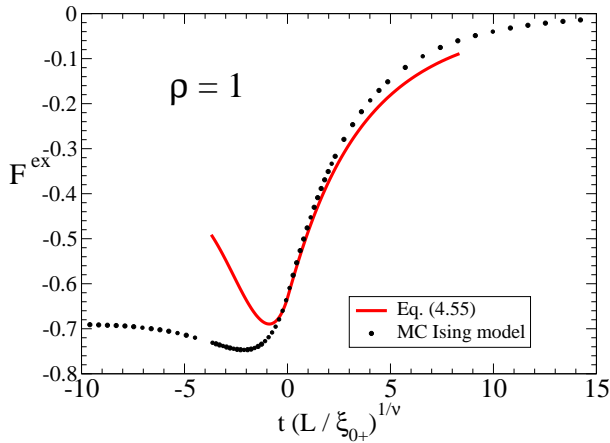}
\caption{(Color online) Scaling function $F^{ex}(\tilde x, 1)$, (\ref{3kk}), (\ref{6l}),  for $n=1$    as a function of  $\tilde x = t (L/\xi_{0+})^{1/\nu}$ in three dimensions
for cubic geometry (solid line) and MC data for the $d=3$ Ising model by Hasenbusch \cite{hasenbusch}. See also Fig. 13 (c).}
\end{figure}
\begin{figure}[!h]
\includegraphics[clip,width=80mm]{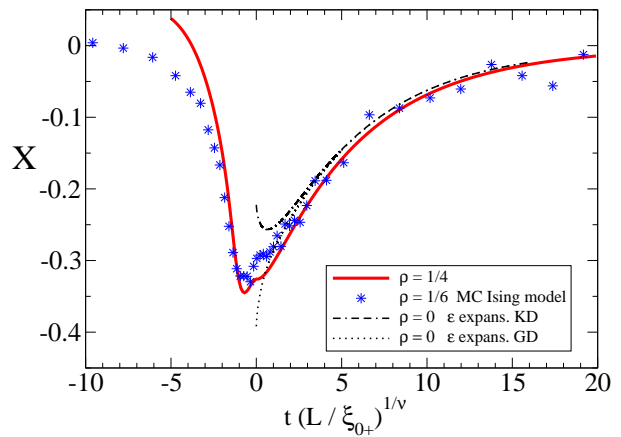}
\caption{(Color online) Scaling function $X(\tilde x, \rho)$, (\ref{3nn}), (\ref{3kk}), (\ref{6l}),  for $n=1$   as a function of  $\tilde x = t (L/\xi_{0+})^{1/\nu}$ in three dimensions
for slab geometry with $\rho=1/4$ (solid line) and MC data for the $d=3$ Ising model with $\rho=1/6$ by Vasilyev et al.  \cite{vasilyev2009}. Thin lines: $\varepsilon$ expansion results for $\rho=0$ from \cite{KrDi92a,GrDi07}. See also Fig. 14 (a).}
\end{figure}

\newpage

\section*{VI. Perturbation theory outside the central finite-size regime}
\renewcommand{\thesection}{\Roman{section}}
\renewcommand{\theequation}{6.\arabic{equation}}
\setcounter{equation}{0}

Outside the central finite-size regime, there is no need for separating the lowest mode, thus ordinary perturbation theory with respect to $u_0$ should be appropriate. By "outside the central finite-size regime" we mean the regions
below the dashed lines in Fig. 2. Within these regions, it is
necessary to further distinguish between scaling and
nonscaling regions (the nonscaling regions correspond to the shaded regions in Fig. 2, see also Fig.1 of \cite{dohm2008}).
The central parts of both the scaling and the nonscaling regions still belong to the asymptotic critical region  $|t|\ll 1$ and $L\gg\tilde a$, $L_{\parallel}\gg\tilde a$.

Here we perform the corresponding analysis  at the
one-loop level. In subsections A - C, we shall consider the scaling region outside the central finite-size regime. The total scaling region can be roughly characterized by  $L/\tilde a\gg 1$, $L_{\parallel}/\tilde a\gg 1$, $\xi_\pm / \tilde a \gg 1$, and $L/\xi_\pm \lesssim 24 (\xi_\pm / \tilde a)^2,
L_\parallel/\xi_\pm \lesssim 24  (\xi_\pm / \tilde a)^2$ where $\xi_\pm$ is the second-moment bulk correlation length above and below $T_c$ , respectively. (Note that this characterization also includes the central finite-size regime which is part of the total scaling region.) The latter restrictions  follow from the conditions (\ref{nonscaling1}) for the nonscaling regions that will be studied in subsection D below. In order to distinguish the perturbation results
of this section from those of Secs. IV and V we use
the notation $f^+_{1-loop}$, $f^-_{1-loop}$, etc.

\subsection*{A. Perturbation theory well above $T_c$}

Ordinary perturbation theory for the excess
free energy density (\ref{2j}) for $n=1$  above $T_c$ yields in one-loop
order
\begin{eqnarray}
\label{VIIa} f^{ex,+}_{1-loop} =   \frac{1}{2 V} \sum_{\bf k} \ln
\{[r_0 - r_{0c} + \delta \widehat K (\mathbf k)] \tilde a^2 \}\nonumber\\- \;\frac{1}{2 }\int\limits_{{\bf k}}  \ln
\{[r_0 - r_{0c} + \delta \widehat K (\mathbf k)] \tilde a^2 \}.\;\;\;
\end{eqnarray}
Here we have already replaced $r_0$ by $r_0-r_{0c}$ which is justified since $r_{0c} \sim O(u_0)$ [see (\ref{4zzz})]. Because of the ${\mathbf k} = {\bf0}$ term, the sum exists only
for $r_0 -r_{0c} > 0$. The evaluation of the excess free energy density
%
%
is outlined in App. A.

In the scaling region  in
$2<d<4$ dimensions, the large-${\bf k}$ dependence of $\delta \widehat K({\bf k})$ does not matter and the leading contribution is obtained by taking  the continuum limit $\tilde a \rightarrow 0$ at
fixed $r_0 - r_{0c} > 0$. For a  renormalization-group (RG) treatment in the scaling region see (10.5) - (10.13) of \cite{dohm2008}. Neglecting nonasymptotic corrections to scaling we obtain the scaling function %
\begin{eqnarray}
\label{VIIkoberhalb}
F^{ex, +}_{1-loop}(\tilde x,\rho) = \frac{1}{2}
\; {\cal G}_0 (\tilde x^{2\nu} ,\rho) \;
\; + \; O (u^*)\;  ,
\end{eqnarray}
where ${\cal G}_0$ is given by (\ref{3.10x}) and $\tilde x^\nu=L/\xi_+$, $\xi_+ = \xi_{0 +} t^{- \nu}$ (for $\xi_{0 +}$ see (5.16) of \cite{dohm2008}). For large  $\tilde x$, $F^{ex, +}_{1-loop}$ decays exponentially to zero according to the asymptotic behavior
\begin{eqnarray}
\label{VIImoberhalb}
&&F^{ex, +}_{asymp}(\tilde x,\rho)  =-
\Big(\frac{\tilde x^\nu}{2\pi}\Big)^{\frac{d-1}{2}} \exp (-\tilde x^\nu
) \nonumber\\&&-\rho^d(d-1)
\Big(\frac{\tilde x_\parallel^\nu}{2\pi}\Big)^{\frac{d-1}{2}} \exp (-\tilde x_\parallel^\nu
) \;\;,
\end{eqnarray}
apart from corrections of $O(e^{-2\tilde x^\nu},e^{-2\tilde x_\parallel^\nu}) $, with  $\tilde x_\parallel^\nu=\tilde x^\nu/\rho$. Eq. (\ref{VIImoberhalb}) follows from (\ref{b30}) in App. A for $\xi_+/\tilde a \gg 1$.  We see that the scaling variable $\tilde x_\parallel$ appears in a natural way in the second term of (\ref{VIImoberhalb})
. For $\rho=1$, (\ref{VIIkoberhalb}) and (\ref{VIImoberhalb}) agree with Eqs. (10.10) - (10.12) of \cite{dohm2008}.

The corresponding scaling functions $\Phi^{ex, +}_{1-loop}$, $X^{+}_{1-loop}$, $\Xi^{ +}_{1-loop}$ and $\Phi^{ex, +}_{asymp}$, $ X^{+}_{asymp}$, $\Xi^{ +}_{asymp}$ follow from (\ref{VIIkoberhalb}), (\ref{VIImoberhalb}), (\ref{3nn}), (\ref{3rr}), and (\ref{3rrr}), respectively.

\subsection*{B. Perturbation theory well below $T_c$}

Perturbation theory for {\it bulk} quantities below $T_c$ within the $\varphi^4$ model for $n=1$ at vanishing external field $h$  may be formulated by first starting with the perturbation expression at finite external field $h>0$ (or $h<0$) and at finite volume $V= L^{d-1}_\parallel L$, then performing the  thermodynamic limit $V \to \infty$ at finite $h>0$ (or $h<0$), and subsequently performing the zero-field limit $h \to 0_+$ (or $h \to 0_-$). Applying this procedure to the free energy density $f(t, L, L_\parallel, h)$ implies that only the contributions of a {\it single} bulk phase with a positive (or negative) spontaneous bulk magnetization are taken into account in the calculation of
\be
\label{fbulk}f_b(t) =\lim_{ h \to 0_+}\lim_{ V\to \infty} f(t, L, L_\parallel, h)  =\lim_{ h \to 0_-}\lim_{  V \to \infty} f(t, L, L_\parallel, h).
\ee
In MC simulations of {\it finite} Ising models at vanishing external field, however, all configurations of both phases with positive and negative magnetization do contribute. In this case, the order-parameter distribution function has two finite peaks with equal heights in the positive and negative ranges of the magnetization \cite{bin-2,Privman-Fisher,binder-landau,cd-1997}. For $T \to 0$, these two peaks are well separated. In order to account for this fact in an analytic treatment of the $\varphi^4$ model well below $T_c$
, it is appropriate to formulate perturbation theory such that an expansion is made around the {\it two} separate peaks of the order-parameter distribution function that exist at $h=0$.
\begin{widetext}
\begin{figure*}[!ht]
\begin{center}
\subfigure{\includegraphics[clip,width=7.92cm]{ExcessfreeEnergyRhoGleichEinviertelAlleTPhysRevE2010.eps}}
\subfigure{\includegraphics[clip,width=7.92cm]{ExcessfreeEnergyRhoGleichEinhalbAlleTPhysRevE2010.eps}}
\subfigure{\includegraphics[clip,width=7.92cm]{ExcessfreeEnergycubeAlleTPhysRevE2010.eps}}
\subfigure{\includegraphics[clip,width=7.92cm]{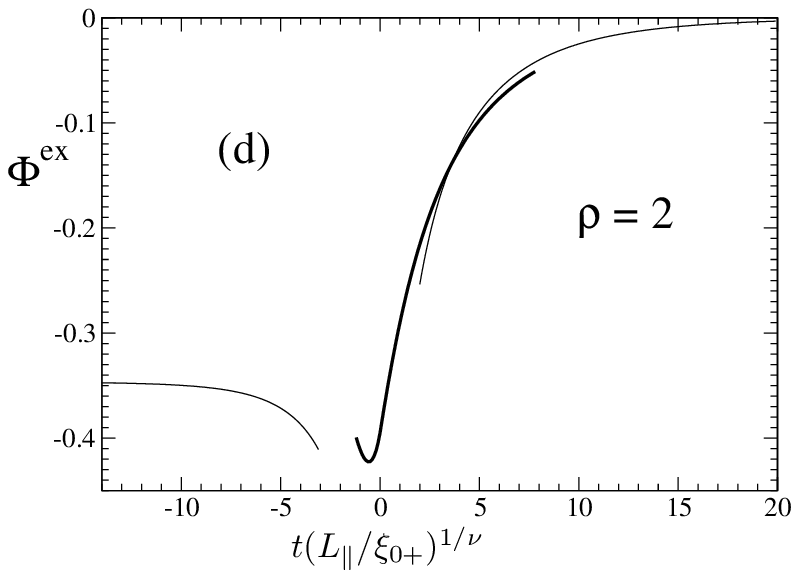}}
\end{center}
\caption{(Color online) Scaling functions $F^{ex}(\tilde x, \rho)$, (a) - (c), and  $\Phi^{ex}(\tilde x_\parallel, \rho)$, (d),   as a function of $\tilde x = t (L/\xi_{0+})^{1/\nu}$ and $\tilde x_\parallel = t (L_\parallel/\xi_{0+})^{1/\nu}$, respectively, for $n=1$ in three dimensions for several values of the aspect ratio $\rho$. Thick lines: improved perturbation theory according to (\ref{6l}) and (\ref{3rr}). Thin lines:  one-loop perturbation theory according to (\ref{VIIkoberhalb}) and (\ref{Fexunterhalb}). The thin lines diverge for $t \to 0$.  The asymptotic value of the thin lines  is $-\rho^2\ln 2$ for $\tilde x \to - \infty$  in (a) - (c) and $-(1/2)\ln 2$ for $\tilde x_\parallel \to - \infty$ in (d). MC data in (c) for  the $d=3$ Ising model by Hasenbusch \cite{hasenbusch}.}
\end{figure*}
\vspace*{0.2cm}

\end{widetext}
\begin{widetext}
\begin{figure*}[!ht]
\begin{center}
\subfigure{\includegraphics[clip,width=7.92cm]{Original-MC-Daten-dgleich3Theorie-Casimir-VonMinus10Bis20PhysRevE2010.eps}}
\subfigure{\includegraphics[clip,width=7.92cm]{Casimir-d=3TheorieRhoGleichEinhalbPhysRevE2010.eps}}
\subfigure{\includegraphics[clip,width=7.92cm]{Casimir-d=3TheorieRhoGleichEinsPhysRevE2010.eps}}
\subfigure{\includegraphics[clip,width=7.92cm]{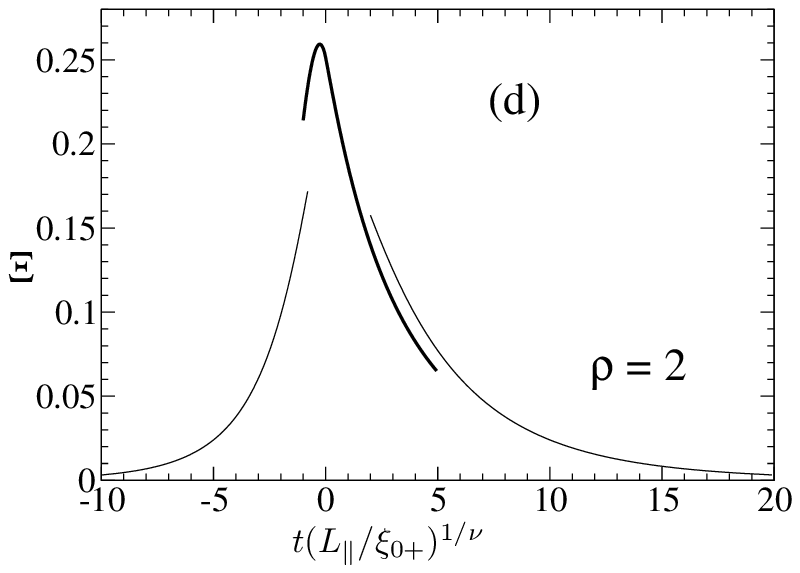}}
\end{center}
\caption{(Color online) Scaling functions $X(\tilde x, \rho)$, (a) - (c), and  $\Xi(\tilde x_\parallel, \rho)$, (d),   as a function of $\tilde x = t (L/\xi_{0+})^{1/\nu}$ and $\tilde x_\parallel = t (L_\parallel/\xi_{0+})^{1/\nu}$, respectively, for $n=1$ in three dimensions for several values of the aspect ratio $\rho$. Thick  lines:  improved perturbation theory according to (\ref{6l}),(\ref{3nn}), and (\ref{3rrr}). Thin lines: one-loop perturbation theory according to (\ref{VIIkoberhalb}) and (\ref{Fexunterhalb}). The thin lines diverge for $t \to 0$. MC data in (a) for  the $d=3$ Ising model with $\rho=1/6$ by Vasilyev et al. \cite{vasilyev2009}.}
\end{figure*}
\vspace*{0.2cm}

\end{widetext}

In the following we perform this approach at the one-loop level in order to calculate $f^{ex,-}_{1-loop}(t,L, L_\parallel, h=0)$ well below $T_c$.
First we  decompose the lattice variable $\varphi_i$ of the Hamiltonian $H$,  (\ref{2a}), as $\varphi_i= M_{>,mf}+ \delta \varphi_i$ with the {\it positive} mean-field order parameter $M_{>, mf} = [-r_0 /(4 u_0)]^{1/2} >0$. Keeping only the Gaussian terms of $H$ up to $O[(\delta \varphi_i)^2]$ corresponding to a one-loop approximation
leads to  the dimensionless partition function  (compare (B1) of \cite{dohm2008})
\begin{eqnarray}
Z_>= \exp\big[V\frac{r_0^2}{16 u_0} \big] \prod_{\bf
k} \left(\frac{2 \pi} {[-2r_0 + \delta \widehat K (\mathbf k)]
\tilde a^2}\right)^{1/2}.
\end{eqnarray}
It can be rewritten as
\begin{eqnarray}
Z_>= \exp(- V f^-_b ) \;Z^{ex}_>
\end{eqnarray}
where $f^-_b$ is the bare bulk free energy density (\ref{4z4}) in one-loop order below $T_c$ and
\be
Z_>^{ex}=\exp \{ V \tilde f^{ex} \}
\ee
is the finite-size part of $Z_>$ with the contribution
\begin{eqnarray}
\tilde f^{ex}=  \;\frac{1}{2 V} \sum_{\bf k} \ln
\{[-2r_0 + \delta \widehat K (\mathbf k)] \tilde a^2 \}\nonumber\\ - \;\frac{1}{2 }\int\limits_{{\bf k}}  \ln
\{[-2r_0 + \delta \widehat K (\mathbf k)] \tilde a^2 \}
\end{eqnarray}
to the excess free energy density below $T_c$. The partition function $Z_>$, however, is incomplete with regard to the finite-size contributions as it does not take into account the fluctuations around the {\it negative} mean-field order parameter $M_{<, mf} = - [-r_0 /(4 u_0)]^{1/2} <0$. A decomposition of $\varphi_i$ as $\varphi_i= M_{<,mf}+ \delta \varphi_i$ 
and an expansion of $H$ up to $O[(\delta \varphi_i)^2]$ leads to the one-loop partition function $Z_<$ which is of course the same as $ Z_>$. Thus the total finite-size part $Z_>^{ex}+Z_<^{ex}$ of the partition function in one-loop order is then given by $ 2 \exp \{ V \tilde f^{ex} \}$. The corresponding total excess free energy density is
\begin{eqnarray}
\label{free-total}
f^{ex,-}_{1-loop} &=& - \frac{\ln 2}{V}\; +\;\frac{1}{2 V} \sum_{\bf k} \ln
\{[-2(r_0-r_{0c}) + \delta \widehat K (\mathbf k)] \tilde a^2 \}\nonumber\\ &-& \;\frac{1}{2 }\int\limits_{{\bf k}}  \ln
\{[-2(r_0-r_{0c}) + \delta \widehat K (\mathbf k)] \tilde a^2 \}   .
\end{eqnarray}
Here we have again replaced $-2 r_0$ by $-2(r_0-r_{0c})$ in the spirit of perturbation theory up to $O(1)$. The result (\ref{free-total}) is identical in form with the (corrected \cite{dohm2010}) result derived previously for cubic geometry \cite{dohm2008}. The non-exponential finite-size term $- V^{-1}\ln 2$  is known from previous work on finite-size effects in Ising models in a block geometry of volume $V$ \cite{Privman-Fisher}. Thus this term is not specific for the $n=1$ $\varphi^4$ theory but rather general for systems with a two-fold degeneracy of the ground state. According to the definition of the Casimir force (\ref{2k}), the constant term $-(\ln 2)/V$ in (\ref{free-total}) does not contribute to $F_{Casimir}$.

The derivation presented above is, of course,  not exact but is valid only well below $T_c$ where the two peaks of the order-parameter distribution function at $h=0$  are well separated and where their wings do not overlap significantly.

The evaluation of (\ref{free-total}) as well as the RG treatment are parallel to that for $f^{ex,+}_{1-loop}$ above $T_c$. Neglecting nonasymptotic corrections to scaling we obtain the scaling function in the scaling region  well below $T_c$
\be
\label{Fexunterhalb}
F^{ex, -}_{1-loop}(\tilde x,  \rho) = - \rho^{d-1}\ln 2\;+\;\frac{1}{2}
\;  {\cal G}_0 ( |2\tilde
x|^{2\nu},\rho) \;
\ee
with
$L/\xi_- =  |2\tilde
x|^\nu$ where
\begin{equation}
\label{VIIp} \xi_- = \xi_{0 -} | t |^{- \nu},\; \xi_{0 - } /
\xi_{0 +} = 2^{- \nu} + O (u^*)
\end{equation}
is the bulk second-moment correlation length  below
$T_c$
and where ${\cal G}_0$ is given by (\ref{3.10x}). In contrast to the vanishing of $F^{ex, +}_{1-loop}$ for $\tilde x \to \infty$, $F^{ex, -}_{1-loop}$ approaches a {\it finite} value $- \rho^{d-1}\ln 2$ for $\tilde x \to -\infty$, as noted already in \cite{dohm2010}. According to (\ref{Fexunterhalb}) and (\ref{VIImoberhalb}), this approach has an exponential form described by the asymptotic behavior
\begin{eqnarray}
\label{VIIm}
&&F^{ex, -}_{asymp}(\tilde x,\rho) = -\rho^{d-1}\ln 2 -
\Big(\frac{|2\tilde x|^\nu}{2\pi}\Big)^{\frac{d-1}{2}} \exp (-|2\tilde x|^\nu
) \nonumber\\&-&\rho^d(d-1)
\Big(\frac{|2\tilde x_\parallel|^\nu}{2\pi}\Big)^{\frac{d-1}{2}} \exp (-|2\tilde x_\parallel|^\nu
)\;\;,
\end{eqnarray}
apart from  corrections of $O(e^{-2|2\tilde x|^\nu},e^{-2|2\tilde x_\parallel|^\nu}) $.

The corresponding scaling functions $\Phi^{ex, -}_{1-loop}$, $X^{-}_{1-loop}$, $\Xi^{ -}_{1-loop}$ and $\Phi^{ex, -}_{asymp}$, $ X^{-}_{asymp}$, $\Xi^{-}_{asymp}$ follow from (\ref{Fexunterhalb}), (\ref{VIIm}), (\ref{3nn}), (\ref{3rr}), and (\ref{3rrr}), respectively.

As noted above, the constant term $-\rho^{d-1} \ln 2$ does not contribute to $X$. This explains why the perturbation result for $X$ of Sec. V (as shown in Fig. 12) is in better agreement with the MC data below $T_c$ than the corresponding result for $F^{ex}$ shown in Fig. 11. Thus, in contrast to $F^{ex, -}$ which has a {\it finite} low-temperature limit $- \rho^{d-1}\ln 2$,  our theory predicts that the Casimir force scaling function has an exponential decay towards {\it zero} for  $\tilde x \to - \infty$. From (\ref{VIIm}) and (\ref{3nn}) we obtain
\begin{eqnarray}
\label{CasExp}
&&X^{-}_{asymp}(\tilde x,\rho) = \nonumber\\&-&\Big(\frac{d-1}{2} + |2\tilde x|^\nu\Big)
\Big(\frac{|2\tilde x|^\nu}{2\pi}\Big)^{\frac{d-1}{2}} \exp (-|2\tilde x|^\nu
) \nonumber\\ &+& \rho^d(d-1)
\Big(\frac{|2\tilde x_\parallel|^\nu}{2\pi}\Big)^{\frac{d-1}{2}} \exp (-|2\tilde x_\parallel|^\nu
).\;\;
\end{eqnarray}

It is suggestive to expect that the formulae (\ref{VIImoberhalb}), (\ref{VIIm}), and (\ref{CasExp}) are
applicable even to $d=2$ - dimensional systems. It would be interesting to check this point for the example of the two-dimensional Ising model with periodic b.c. in rectangular geometry.

\subsection*{C. Predictions for the whole scaling region }

On the basis of the three perturbation results (\ref{6l}), (\ref{VIIkoberhalb}), and (\ref{Fexunterhalb}) we are now in the position  to present quantitative predictions  for the various scaling functions over the whole range of the scaling variables $-15 \lesssim \tilde x \lesssim 20$ and $-15 \lesssim \tilde x_\parallel \lesssim 20$. These scaling functions are shown in Figs. 13 and 14 for various values of the aspect ratio $\rho$ in three dimensions.

The thin lines  are based on one-loop perturbation theory (\ref{VIIkoberhalb}) and (\ref{Fexunterhalb}) and are applicable only away from $T_c$ outside the central finite-size regime. For $T \to T_c$, one-loop perturbation theory breaks down which implies  that the thin lines diverge for $t \to 0$.  The thick lines  are based on our lowest-mode separation approach presented in Secs. IV and V which is applicable to the central finite-size regime including $T=T_c$. This improved perturbation approach provides a bridge through $T=T_c$ between the simple finite-size critical behavior represented by the thin lines well away from $T_c$. The lowest-mode separation approach is not applicable, however, to the regions $|\tilde x| \gg 1$ and $|\tilde x_\parallel| \gg 1$. Our Figs. 13 and 14 demonstrate that one-loop perturbation theory and improved perturbation theory complement each other and match reasonably well at intermediate values of the scaling variables. No perfect matching can be
expected because of missing $O(u^*)$ terms in the one-loop results. Comparison with the MC data in Figs. 13 and 14  shows that the improvement achieved by the one-loop results is clearly visible in the range $\tilde x > 4$ and $\tilde x < -1$ [in Fig. 13 (c)] and in the range $\tilde x < -2$ [in Fig. 14 (a)]. On the whole, we consider the good agreement of our theory with the MC data  over the entire scaling regime $-15 \lesssim \tilde x \lesssim 20$ as a major success of our strategy employing three different perturbation approaches. Comparison with  MC data for other values of $\rho$ would be interesting.

\subsection*{D. Exponential nonscaling region}

So far we have eliminated the dependence on the lattice spacing $\tilde a$ by taking the continuum limit.  In earlier work
it was pointed out for confined systems in an $L^d$ geometry \cite{dohm2008,cd2000-1} and in film geometry \cite{kastening-dohm} that the finite lattice constant $\tilde a$ becomes
non-negligible in the limit of large $L/\tilde a$ at fixed $T\neq
T_c $ in the regime where the finite-size scaling function has an
exponential form. The same arguments apply to the present system in a finite block geometry.
As shown in
App. A, the excess free energy density in one-loop order attains
the following form in the limit of large $L/\tilde a  $, large $L_\parallel/\tilde a$, large $L/\xi_\pm $, and large $L_\parallel/\xi_\pm$

\begin{eqnarray}
\label{fasympplus} f^{ex, +}_{asymp}= {\cal A}_+(\xi_+,L,L_\parallel,\tilde a),
\end{eqnarray}
\begin{eqnarray}
\label{fasympminus} f^{ex, -}_{asymp}= -\frac{\ln 2}{V} + {\cal A}_-(\xi_-,L,L_\parallel,\tilde a),
\end{eqnarray}
with the nonuniversal function
\begin{eqnarray}
\label{VIIoneuneu} &&{\cal A}_\pm(\xi_\pm,L,L_\parallel,\tilde a)= \nonumber\\&- &\frac{1}{L^d} \left[1 + \left(\frac{\tilde a}{2
\xi_\pm}\right)^2 \right]^{\frac{d-1}{4}}
\left(\frac{L}{2\pi \xi_\pm}\right)^{\frac{d-1}{2}} \exp \left\{-
\frac{L}{\xi_{{\bf e} \pm} }  \right\} \;
\nonumber\\ &-&\frac{d-1}{L_\parallel^d} \left[1 + \left(\frac{\tilde a}{2
\xi_\pm}\right)^2 \right]^{\frac{d-1}{4}}
\left(\frac{L_\parallel}{2\pi \xi_\pm}\right)^{\frac{d-1}{2}} \exp \left\{-
\frac{L_\parallel}{\xi_{{\bf e} \pm} }  \right\},\nonumber\\
\end{eqnarray}
where
\begin{eqnarray}
\label{VIIr} \xi_{{\bf e} \pm} \; = \; \frac{\tilde a}{2}
\left[{\rm arsinh} \left(\frac{\tilde a}{2
\xi_\pm}\right)\right]^{-1}
\end{eqnarray}
are the {\it exponential ("true")} bulk correlation lengths \cite{dohm2008,cd2000-2,fish-2} above (+) and below (-) $T_c$, respectively.
This result applies to the regions well below the dashed lines in Fig. 2 including the shaded regions. Note that no condition is imposed on the value of $0<\tilde a/
\xi_\pm < \infty$ other than that $L/\xi_\pm $ and $L_\parallel/\xi_\pm$ are large. For $L=L_\parallel$, (\ref{VIIoneuneu}) reduces to the previous result for cubic geometry \cite{dohm2008,footnote}.
As a nontrivial relation
between bulk properties and finite-size effects \cite{cd2000-2},
the lengths $\xi_{{\bf e}\pm}$ describe the exponential part of
the {\it bulk} order-parameter correlation function \cite{fish-2}
in the large-distance limit in the direction of one of the cubic
axes at arbitrary fixed $T \neq T_c$ above and below $T_c$ (for $n
= 1$), respectively. This relation is exact in the large-$n$ limit
above $T_c$ \cite{cd2000-2}.

It has been shown \cite{dohm2008,cd2000-1} that, because of the exponential
structure of the function ${\cal A}_\pm$,  the
$\tilde a$ dependence of $\xi_{{\bf e}\pm}$ cannot be neglected
even for small $\tilde a / \xi_\pm \ll 1$ if
\begin{eqnarray}
\label{nonscaling1}L \gtrsim  24  \xi_\pm^3 / \tilde a^2, \;\;L_\parallel \gtrsim  24  \xi_\pm^3 / \tilde a^2
\end{eqnarray}
are sufficiently large. The conditions (\ref{nonscaling1}) follow from the second term in the expansion of the function (\ref{VIIr}) for small $\tilde a / \xi_\pm $
\begin{eqnarray}
\label{expand} \xi_{{\bf e} \pm} \;= \xi_\pm \Bigg[1 - \frac{1}{24}\Big(\frac{\tilde a}{\xi_\pm}\Big)^2 + \cdot\cdot\cdot \Bigg]
\end{eqnarray}
appearing in the exponential parts of the function ${\cal A}_\pm(\xi_\pm,L,L_\parallel,\tilde a)$ (see also  \cite{cd2000-1,cd2000-2}). The second term in (\ref{expand}) is {\it not} negligible even for small $\tilde a / \xi_\pm \ll 1$ if the conditions (\ref{nonscaling1})  are satisfied.
This implies that finite-size scaling and universality are
violated in the large $ |\tilde x| $ and $ |\tilde x_\parallel| $ tail of $L^d f_s^{ex}$ at any $\tilde a / \xi_\pm > 0$ even arbitrarily close to
$T_c$ because ultimately, for $|\tilde x| \to \infty$ and $|\tilde x_\parallel| \to \infty$ (i.e., for large
$L$ and $L_\parallel$ at fixed $|t| > 0$), the tail of $L^d f_s^{ex}$
becomes explicitly dependent on $\tilde a$. As shown in Sec. X of \cite{dohm2008}, the tail depends even on the bare four-point coupling $u_0$ through $\xi_\pm $: strictly speaking it is even necessary to keep the complete nonasymptotic ($u_0$ dependent) form of $\xi_\pm$ at finite $\tilde a$.
Thus no $\tilde a$ - independent finite-scaling form
with a single scaling argument $\propto t L^{1/\nu}$ can be defined in this exponential large
$ |\tilde x| $ and large $ |\tilde x_\parallel| $ region.  Higher-loop contributions
cannot remedy this violation.
The same reservations apply, of course, to the critical Casimir force and its scaling form.

{\it Note added}: The predictions of Ref. \cite{dohm2010} and of the present paper are in good agreement with recent Monte Carlo data for the three-dimensional Ising model by A. Hucht, D. Gr\"{u}neberg, and F.M. Schmidt, Phys. Rev. E {\bf 83}, 051101 (2011).

\renewcommand{\thesection}{\Roman{section}}
\setcounter{equation}{0} \setcounter{section}{1}
\renewcommand{\theequation}{\Alph{section}.\arabic{equation}}
\section*{ACKNOWLEDGMENT}

I am grateful to M. Hasenbusch for providing the MC data of Ref. \cite{hasenbusch} in numerical form prior to publication. I also thank A. Hucht and  B. Kastening  for useful discussions and correspondence.

\renewcommand{\thesection}{\Roman{section}}
\setcounter{equation}{0} \setcounter{section}{1}
\renewcommand{\theequation}{\Alph{section}.\arabic{equation}}
\section*{Appendix A: Gaussian free energy}

We consider the Gaussian model, i.e., the Hamiltonian (\ref{2a}) for $u_0=0$, and calculate the excess free energy density  in
a rectangular$L_\parallel^{d-1} \times L$ geometry. This calculation will  lead to the  evaluation of the sums in  (\ref{3a}) and (\ref{4b}) as well as to the derivation of (\ref{4rrp0}) - (\ref{4.28}), (\ref{VIIkoberhalb}), (\ref{VIImoberhalb}), (\ref{Fexunterhalb}), (\ref{VIIm}), and (\ref{VIIoneuneu}). Since the calculation is largely parallel to that of \cite{dohm2008} we skip some of the details of the derivation.

The Gaussian excess free energy density per component divided by $k_BT$ is
$ f^{ex}_{Gauss}= \frac{1}{2}\Delta(r_0, L_\parallel, L, K_{i,j}, \tilde a)$,
\begin{eqnarray}
\label{b9} &&\Delta(r_0, L_\parallel, L, K_{i,j}, \tilde a) = V^{-1}
{\sum_{\bf k}} \ln
\{[r_0 + \delta \widehat K (\mathbf k)] \tilde a^2\}\nonumber\\
&-& \int\limits_{\bf k} \ln \{[r_0 + \delta \widehat K (\mathbf k)]
\tilde a^2\}
\end{eqnarray}
where the sum $\sum_{\bf k}$ and the integral  $\int_{\bf k}$ have
finite cutoffs $\pm \pi / \tilde a$ for each $k_\alpha$.
Using the Poisson identity
\cite{cd2000-2,morse} we obtain the exact representation
\begin{eqnarray}
\label{b13} \Delta(r_0,  L_\parallel, L, K_{i,j}, \tilde a) \; = \;-\int
\limits_0^\infty dy y^{-1} e^{-r_0 \tilde a^2 y} \nonumber \\
\times {\sum_{{\bf m}, n}}' \int\limits_{\bf q} \int\limits_p \exp
\{ -  \delta \widehat K (\mathbf q, p) \tilde a^2 y + i {\bf q}
\cdot {\bf m}  L_\parallel\ + i p
 n  L\}\;\;\;
\end{eqnarray}
with ${\bf q} \cdot {\bf
m} = \sum^{d-1}_{\alpha = 1} q_\alpha m_\alpha$ where ${\sum'_{{\bf m}, n}}$ means summation over all integers ${\bf m}= (m_1, m_2, ..., m_{d-1})$ and $n$ without the single term with ${\bf m =0 }, n=0$.
In the following we evaluate $\Delta$
for $L \gg \tilde a$ and $L_\parallel \gg \tilde a$ in two regimes.

\subsection*{1. Central finite-size regime}

We assume  large
$L/\tilde a$, large $ L_\parallel/\tilde a$, small $0 < r_0^{1/2}\tilde a \ll 1$ and fixed $ 0 < Lr_0^{1/2}\lesssim O(1)$, $ 0 < L_\parallel r_0^{1/2}\lesssim O(1)$ which we refer to as the central finite-size regime. In this regime, the large - (${\bf q}, p$) dependence of $\delta
\widehat K (\mathbf q, p)$ does not matter. Therefore we may replace
$\delta \widehat K (\mathbf q, p)$ by its long - wavelength form (\ref{2h})
and let the integration limits of
$\int_{\bf q}$ and of $\int_p$ go to $\infty$. This leads to the scaling form of the Gaussian excess free energy
\be
\label{F.15}
f^{ex}_{Gauss}=\frac{1}{2} \Delta (r_0, L_{\parallel}, L ) =\frac{1}{2}  L^{-d}{\cal G}_0
(r_0 L^2, \rho)
\ee
where ${\cal G}_j (r_0 L^2, \rho)$ is defined in  (\ref{3.10x}). Interpreting (\ref{F.15}) as a one-loop contribution of the $\varphi^4$ model and applying the renormalization procedure parallel to that described in Sec. X A. of \cite{dohm2008} we arrive at the one-loop scaling function presented in (\ref{VIIkoberhalb}).

The function ${\cal G}_0 (r_0 L^2, \rho)$ diverges for $r_0 L^2 \to 0$ which comes from the large-$z$ behavior of  $K(z) \approx 1$ in the last term $\Big[\rho K(\rho^2z)\Big]^{d-1}K(z) \approx \rho^{d-1}$ of the integrand of (\ref{3.10x}). We find
\begin{eqnarray}
\label{Gnullasym}
{\cal G}_0 (r_0 L^2, \rho) \approx \rho^{d-1}\ln \Big(\frac{r_0 L^2}{4\pi^2}\Big) + {\cal C}_0(\rho)
\end{eqnarray}
for  $r_0 L^2 \ll 1$. In order to determine the constant ${\cal C}_0(\rho)$ we add and subtract the divergent term $\rho^{d-1}\ln [(r_0 L^2/(4\pi^2)]$ by rewriting  ${\cal G}_0 (r_0 L^2, \rho)$ in the form
\begin{eqnarray}
\label{F.15x} {\cal G}_0(r_0 L^2, \rho)= \rho^{d-1}\ln \Big(\frac{r_0 L^2}{4\pi^2}\Big)+\int\limits_0^\infty \frac{dz}{z}\Bigg[
  \exp {\left(-\frac{r_0 L^2z}{4\pi^2}\right)}
  \nonumber\\ \times \left\{\Big(\frac{\pi}{z}\Big)^{d/2}-\Big[\rho K(\rho^2z)\Big]^{d-1}K(z)
    + \rho^{d-1}\right\} -\rho^{d-1} e^{-z}\Bigg].\nonumber\\
\end{eqnarray}
The integral in (\ref{F.15x}) has a finite limit for $r_0 L^2 \to 0$ which yields the constant
\begin{eqnarray}
\label{F.15xx} &&{\cal C}_0(\rho)= \int\limits_0^\infty \frac{dz}{z}  \Bigg[
\Big(\frac{\pi}{z}\Big)^{d/2}\nonumber\\&-&\Big[\rho K(\rho^2z)\Big]^{d-1}K(z)
    + \rho^{d-1}(1 -e^{-z})\Bigg].
\end{eqnarray}
Eq. (\ref{Gnullasym}) implies that the function
\begin{eqnarray}
\label{3.10xxx} {\cal G}_1(r_0 L^2, \rho)=-\frac{\partial {\cal G}_0(r_0 L^2, \rho)}{\partial(r_0L^2)}
\end{eqnarray}
has the divergent behavior
\begin{eqnarray}
\label{Geinsasym}
{\cal G}_1 (r_0 L^2, \rho) \approx -\frac{\rho^{d-1}}{r_0 L^2}
\end{eqnarray}
for  $r_0 L^2 \ll 1$. The asymptotic behavior (\ref{Gnullasym}) and (\ref{Geinsasym}) is needed in the discussion of the low-temperature limit in Sec. III.C .

The function ${\cal G}_0 (r_0 L^2, \rho)$ decays exponentially for large $r_0 L^2 \gg 1$. From (\ref{b30}) we obtain for $r_0\tilde a^2 \ll1$ and  $r_0 L^2 \gg 1$
\begin{eqnarray}
\label{Gnullasymplus} {\cal G}_0 (r_0 L^2, \rho) \approx  - 2
\left(\frac{r_0 L^2}{4 \pi^2 }\right)^{(d-1)/4} \exp(-
Lr_0^{1/2})\nonumber\\  - 2(d-1) \rho^d
\left(\frac{r_0 L^2}{4 \pi^2 \rho^2}\right)^{(d-1)/4} \exp(-L r_0^{1/2}/\rho
) .
\end{eqnarray}
For $\rho=1$, Eq. (\ref{Gnullasymplus}) agrees with Eq. (10.12) of \cite{dohm2008}.

The result (\ref{F.15}) is sufficient to derive the higher-mode sums  $S_i(r_{0{\rm L}}, L, \rho)$, (\ref{S0}) and (\ref{4rr}) in the central finite-size regime. We obtain
\begin{eqnarray}
\label{c111}&&\frac{1}{V} {\sum_{\bf k\neq0}} \ln \{[r_0 + \delta
\widehat K (\mathbf k)] {\tilde a}^2\} \nonumber\\& = &
\int\limits_{\bf k} \ln \{[r_0 + \delta \widehat K
(\mathbf k)] {\tilde a}^2\}+ \frac{1}{L^d}\ln
\left(\frac{L^2}{{\tilde a}^2 4\pi^2}\right) \nonumber\\&+& \frac{1-\rho^{d-1}}{L^d} \ln(r_0  {\tilde a}^2)   + \;\frac{1}{L^d}
J_0(r_0 L^2, \rho)
\end{eqnarray}
with $ J_0(x^2, \rho)$ defined by (\ref{4.27}).
By means of differentiation with
respect to $r_0$ we obtain from (\ref{c111}) for $m=1,2$
\begin{eqnarray}
\label{c33}&&V^{-1} \sum_{{\bf k} \neq {\bf 0}} [r_0 +
\delta \widehat K(\mathbf k)] ^{-m} =
\int\limits_{\bf k}[r_0 + \delta \widehat K (\mathbf k)]^{-m}\nonumber\\& +&
\frac{1-\rho^{d-1}}{L^d} (r_0)^{-m}+ \frac{L^{2m-d}}{(4 \pi^2)^m}
I_m (r_0L^2, \rho)
\end{eqnarray}
with  $I_m (r_0L^2, \rho)$ defined by (\ref{4.28}). For  the bulk integrals see \cite{dohm2008}.

\subsection*{2. Exponential regime above $T_c$}
Now we assume $L r_0^{1/2} \gg 1$ and $L_{\parallel} r_0^{1/2} \gg 1$  at finite $\rho=L/L_{\parallel}$ for fixed $r_0^{1/2} \tilde a > 0$ which we refer to as the exponential regime since $\Delta(r_0,  L_\parallel, L, K_{i,j}, \tilde a)$ will
attain an exponential $L$ and  $L_\parallel$ dependence in this regime. In this regime the complete ${\bf k}$
dependence of the microscopic interaction $\delta \widehat K (\mathbf k)$ does matter.
We use the nearest-neighbor interaction (\ref{2g}) in the form
\begin{eqnarray}
 \delta \widehat K ({\mathbf q}, p) =  \frac{2}{\tilde a^2} \sum_{\alpha = 1}^{d-1}  \left[1 - \cos (\tilde a
q_\alpha) \right] +  \frac{2}{\tilde a^2} \left[1 - \cos (\tilde
a p) \right].\;\;\;
\end{eqnarray}
Generalizing the derivation of \cite{dohm2008} to block geometry we obtain
for large $L/\tilde a$ and  $L_\parallel/\tilde a$ but arbitrary $\tilde r_0 \equiv r_0 \tilde a^2 > 0$ (compatible with $L r_0^{1/2} \gg 1$ and $L_{\parallel} r_0^{1/2} \gg 1$)
\begin{eqnarray}
\label{b40} &&\Delta (r_0, L_\parallel, L, K_{i,j}, \tilde a) = \nonumber\\&-&\frac{ 2}{\tilde a^{d}(2\pi L/\tilde a)^{d/2}
} \int\limits_0^\infty dz  \;\frac{ \exp [\Phi(z,\tilde r_0)L/\tilde a ]}{z^{(d+1)/2}q^{1/2}}
\nonumber\\ &-&\frac{ 2(d-1)}{\tilde a^{d}(2\pi L_\parallel/\tilde a)^{d/2}
} \int\limits_0^\infty dz  \;\frac{ \exp [\Phi(z,\tilde r_0)L_\parallel/\tilde a] }{z^{(d+1)/2}q^{1/2}}
\end{eqnarray}
where $q = (1+z^2)^{1/2}$ with
\begin{eqnarray}
\label{b31} \Phi(z,\tilde r_0)= - (1+\tilde r_0/2)z + q +\ln \Big(\frac{z}{1+q} \Big).
\end{eqnarray}
The maximum of the function $\Phi(z,\tilde r_0)$ in the
exponential parts of the integrand of (\ref{b40}) is at $z = \bar
z$ where
\be
\bar z \; = \; \left[\tilde r_0 \Big(1 + \frac{\tilde r_0}{
4}\Big)\right]^{- 1/2}.
\ee
Expanding $\Phi(z,\tilde r_0)$ around $z = \bar z$ up to $O[(z-\bar z)^2]$ and performing
the integration over $z$ we finally obtain the Gaussian excess free energy density for large $L/\tilde a$ and large $L_\parallel/\tilde a$ at
arbitrary fixed $r_0  > 0$
\begin{eqnarray}
\label{b30} f^{ex}_{Gauss,asymp}=  - \frac{1}{L^d}
\left(\frac{L/\tilde a}{2 \pi \bar z}\right)^{(d-1)/2} e^{-
L/\xi_{\bf e}^G}\nonumber\\  - \frac{(d-1)}{L_\parallel^d}
\left(\frac{L_\parallel/\tilde a}{2 \pi \bar z}\right)^{(d-1)/2} e^{-
L_\parallel/\xi_{\bf e}^G}
\end{eqnarray}
with the {\it  exponential} ("true") bulk correlation length of the Gaussian model
\begin{eqnarray}
\label{b41} \xi_{\bf e}^G \; = \; \frac{\tilde a}{2} \left[{\rm
arsinh} \left(\frac{r_0^{1/2}\tilde a }{2 }\right)\right]^{-1} \; .
\end{eqnarray}
We recall that $r_0^{-1/2}=\xi_+^G$ is the {\it second-moment} bulk correlation length of the Gaussian model above $T_c$. For $L=L_\parallel$ (cube), (\ref{b30}) yields the previous result of Eq. (B24) of \cite{dohm2008}. No universal finite-size scaling function of the Gaussian model
can be defined in the region $L \gtrsim  24  (\xi_+^G)^3 / \tilde a^2$ and $L_\parallel \gtrsim  24  (\xi_+^G)^3 / \tilde a^2$  because of the
explicit $\tilde a$ dependence of (\ref{b30}) and (\ref{b41}).

Within a RG treatment of the $ \varphi^4$ lattice model the
Gaussian results (\ref{F.15}) and (\ref{b30})  can be considered as the bare one-loop
contributions to the excess free energy density. By means of such a RG treatment at finite lattice
constant $\tilde a$ parallel to Sect. 2 and App. A of
\cite{cd2000-1}, these results  acquire the
correct critical exponents of the $n=1$ universality class
including corrections to scaling. This leads to the one-loop
results at finite $\tilde a$ in Sec. VI D.


\begin{thebibliography}{99}
\bibitem{film}
In Fig. 15 of \cite{vasilyev2009}, the location of the film transition for the 3 $d$ Ising model  with isotropic interactions is indicated at the value $x^* = - 1.6$ of the scaling variable $x$, slightly below the minimum. Note that the value of $x^*$ is universal only for isotropic systems  whereas it is expected to be nonuniversal for anisotropic systems  \cite{kastening-dohm}.
\end{thebibliography}
\end{document}